\documentclass[usenatbib]{mn2e}
\usepackage{times,graphicx,url}
\usepackage{amssymb,amsmath}
\usepackage{aas_macros}
\usepackage{subfigure}
\usepackage{ulem}

\newcommand{\besancon}{Besan{\c c}on}
\newcommand{\euclid}{{\it Euclid}}
\newcommand{\exels}{ExELS}
\newcommand{\wfirst}{{\it WFIRST}}
\newcommand{\kepler}{{\it Kepler}}
\newcommand{\dune}{{\it DUNE}}
\newcommand{\harps}{{\it HARPS}}

\newcommand{\mabuls}{{\sc mab$\mu$ls}}
\newcommand{\vis}{{\sc vis}}
\newcommand{\nisp}{{\sc nisp}}

\newcommand{\thetae}{\theta_{\mathrm{E}}}
\newcommand{\re}{r_{\mathrm{E}}}

\newcommand{\dl}{D_{\mathrm{l}}}
\newcommand{\ds}{D_{\mathrm{s}}}

\newcommand{\tein}{t_{\mathrm{E}}}
\newcommand{\murel}{\mu_{\mathrm{rel}}}
\newcommand{\tzero}{t_{\mathrm{0}}}
\newcommand{\uzero}{u_{\mathrm{0}}}
\newcommand{\rhostar}{\rho}

\newcommand{\msun}{M_{\odot}}
\newcommand{\mearth}{M_{\earth}}
\newcommand{\rearth}{R_{\earth}}

\newcommand{\dd}{\mathrm{d}}

\newcommand{\kms}{km~s$^{-1}$}

\newcommand{\mpl}{M_{\mathrm{p}}}
\newcommand{\mffp}{M_{\mathrm{ffp}}}
\newcommand{\rpl}{R_{\mathrm{p}}}

\newcommand{\rtau}{f_{\Gamma}}

\title[\exels{}: an exoplanets survey with Euclid I.]{\exels{}: an exoplanet legacy science proposal for the ESA \euclid{} mission I. Cold exoplanets}

\author[M.T.~Penny et al.]{M.T.~Penny,$^{1,2,3}$ E.~Kerins,$^{1,2}$\thanks{Correspondence to: Eamonn.Kerins@manchester.ac.uk}
  N.~Rattenbury,$^2$ J.-P.~Beaulieu,$^{1,4}$ A.C.~Robin,$^5$ S.~Mao,$^{1,2,6}$
\newauthor
  V.~Batista,$^{1,3}$ S.~Calchi~Novati,$^{1,7,8}$ A.~Cassan,$^{1,4}$ P.~Fouqu\'{e},$^{1,9}$  I.~McDonald,$^{1,2}$
\newauthor
  J.B.~Marquette,$^{1,4}$ P.~Tisserand,$^{1,10}$ M.R.~Zapatero~Osorio$^{1,11}$\\
$^1$ The Euclid Exoplanet Science Working Group\\
$^2$Jodrell Bank Centre for Astrophysics, School of Physics \&
Astronomy, University of Manchester, Oxford Road, Manchester M13 9PL, UK\\
$^3$Department of Astronomy, Ohio State University, 140 W. 18th Ave., Columbus, OH 43210, USA\\
$^4$Institut d`Astrophysique de Paris, Universit\'{e} Pierre et Marie Curie, CNRS UMR7095, 98bis Boulevard Arago, 75014 Paris, France\\
$^5$Institut Utinam, CNRS UMR6213, Universit\'{e} de
Franche-Comt\'{e}, Observatoire de Besan\c{c}on, Besan\c{c}on,
France\\
$^6$National Astronomical Observatories, Chinese Academy of Sciences, A20 Datun
Road, Chaoyang District, Beijing 100012, China\\
$^7$Dipartimento di Fisica ``E. R. Caianiello'', Universit\`a di 
Salerno, Via Ponte don Melillo, 84084 Fisciano (SA), Italy\\
$^8$Istituto Internazionale per gli Alti Studi Scientifici (IIASS), 
Vietri Sul Mare (SA), Italy\\
$^9$IRAP, CNRS - Université de Toulouse, 14 av. E. Belin, F-31400 Toulouse, France \\
$^{10}$Research School of Astronomy and Astrophysics, Australian National University, Cotter Rd, Weston Creek, ACT, 2611, Australia\\
$^{11}$Centro de Astrobiolog\'{i}a (CSIC-INTA), Crta. Ajalvir km 4, E-28850 Torrej\'{o}n de Ardoz, Madrid, Spain
}

\begin{document}

\newcommand{\be}{\begin{equation}}
\newcommand{\ee}{\end{equation}}
\newcommand{\umax}{u_{\mathrm{0max}}}
\newcommand{\fpivot}{f_{\bullet}}
\newcommand{\mpivot}{M_{\bullet}}
\newcommand{\icarus}{{\it Icar.}}

\maketitle

\begin{abstract}
The \euclid{} mission is the second M-class mission of the ESA Cosmic Vision programme, with 
the principal science goal of studying dark
energy 
through observations of weak lensing and baryon acoustic oscillations. \euclid{} is also expected to undertake
additional Legacy Science programmes. One such proposal is the Exoplanet Euclid Legacy Survey (\exels{}) which will be the first survey able to measure the abundance of exoplanets down to Earth mass for host separations from $\sim$1~AU out to the free-floating (unbound) regime. The cold and free-floating exoplanet regimes represent a crucial discovery space for testing planet formation theories. \exels{} will use the gravitational microlensing technique and will detect 1000 microlensing events per month over 1.6~deg$^2$ of the Galactic bulge. We assess how many of these events will have detectable planetary signatures using a detailed multi-wavelength microlensing simulator --- the Manchester-\besancon{} microLensing Simulator (\mabuls{}) --- which incorporates the
\besancon{} Galactic model with 3D extinction. \mabuls{} is the first theoretical simulation of microlensing to treat the effects of point spread function (PSF) blending self-consistently with the underlying Galactic model. We use \mabuls{}, together with current numerical models for the \euclid{} PSFs, to explore a number of designs and de-scope options for \exels{}, including the exoplanet yield as a function of filter choice and slewing time, and the effect
of systematic photometry errors. Using conservative extrapolations of current empirical exoplanet mass functions determined from ground-based microlensing and radial velocity surveys, \exels{} can expect to
detect a few hundred cold exoplanets around mainly G, K and M-type stellar hosts, including ${\sim}45$ Earth-mass planets and ${\sim}6$ Mars-mass planets for an observing programme totalling 10 months. \exels{} will be capable of measuring the cold exoplanet mass function down
to Earth mass or below, with orbital separations ranging from
${\sim}1$~AU out to infinity (i.e. the free-floating regime). Recent ground-based microlensing measurements indicate a significant population of free-floating Jupiters, in which case \exels{} will detect hundreds of free-floating planets. \exels{} will also be sensitive to hot exoplanets and sub-stellar companions through their transit signatures and this is explored in a companion paper. 
\end{abstract}


\newcommand{\figuresize}{84mm}

\section{Introduction} \label{intro}

The discovery of exoplanetary systems is accelerating rapidly with over 860 exoplanets confirmed from ground- and space-based observations\footnote{As of April 2013. See the Extrasolar Planets Encylopaedia: \url{http://exoplanet.eu/}} and another ${\sim}2700$ candidates detected with the \kepler{} space telescope \citep{Batalha:2013}. This is providing a wealth of knowledge on the distribution function of, primarily, hot exoplanets at host separations $\la 1$~AU around FGK-type stars. Recent observations by the \kepler{} space-based transit mission indicate that low-mass exoplanets appear to be common and that around 20\% of stellar hosts have multiple planets orbiting them \citep{Batalha:2013}. Results from 8 years of observations by the \harps{} radial velocity team \citep{Mayor:2011hor} indicate that half of Solar-type stars host planets with orbital periods below 100 days. The frequency of exoplanets in the Super-Earth to Neptune (SEN) mass range shows a sharp increase with declining mass and no preference for host star metallicity. \harps{} also finds that most SEN planets belong to multiple exoplanet systems. 
An analysis by \harps{} of its M~dwarf star sample \citep{Bonfils:2013} indicates that low-mass exoplanets are also common around low-mass stars and that the fraction $\eta$ of M~dwarf host stars with habitable planets is remarkably high at $\eta = 0.41^{+0.54}_{-0.13}$.

The vast majority of low-mass exoplanet detections to date are ``hot'', involving planets within $\sim 1$~AU of their host star. Currently only 8 ``cool'' exoplanets have been detected with masses below 30~$M_{\oplus}$ and host separations above 1~AU. This reflects the fact that such exoplanets are highly demanding targets for both the transit and radial velocity detection methods, techniques which dominate current exoplanet statistics. 

Mapping the cold exoplanet regime is crucial for testing and informing leading theories of planet formation, such as the core accretion and disk instability scenarios.
In the core accretion scenario~\citep{Safronov:1969cam,
  Mizuno:1980cam, Lissauer:1987cam}, planets form out of a thick disc
of gas and dust by the gradual build-up of material from dust grains
into larger objects through collisions. Once the objects
become large enough, they begin to accrete dust and gas
via gravity. In the core accretion model,
terrestrial planets can be considered as the
cores of planets that fail to reach the mass required for runaway
gas accretion, either due to their location in the disc or the
influence of other planets nearby that grow more rapidly. The core
accretion process is most efficient in a region of enhanced disc
density where water and other hydrogen compounds condense to form ice~\citep{Hayashi:1981, Stevenson:1988sno}. This region (the so-called ice- or snow-line)
lies at orbital radii ${\sim} 2.7$~AU, likely with only a weak host-mass dependence, and is thought to be where most
planets form. 
In the disc instability scenario~\citep{Kuiper:1951oss,
  Cameron:1978pgi, Boss:1997pgi}, giant planets form through a
gravitational instability in a gaseous disc. Disc instability may be the only mechanism by which giant
planets can form~\citep{Boss:2011dif}, whilst terrestrial
planets are still thought to form through a process similar to core
accretion~\citep{Boss:2006sei}.
Migration of planets during
formation, due to interactions with the disc, can cause both
inward~\citep{Goldreich:1980pfm, Ward:1997tot} and
outward migration~\citep{Masset:2001opm}. More violent planet-planet
interactions may result in planets being scattered inwards
\citep{Nagasawa:2008pps}, outwards or even being ejected completely
from their systems~\citep{Veras:2009pps}. Tentative evidence of
unbound (free-floating) planetary-mass
objects suggest that more than one Jupiter-mass planet per star may
be ejected in this way~\citep{Sumi:2011ffp}.

Of the relatively few cool low-mass exoplanets detected to date at host separations above 1~AU and mass below 30 Earth masses, half have been found using the gravitational microlensing technique \citep{Mao:1991bml,Gould:1992pmm}. The peak sensitivity of microlensing occurs around the location of the snow line, making it a particularly powerful probe of planet formation. It is also sensitive to free-floating planets whose existence may provide an additional `smoking-gun' signature of the planet formation process. 

Whilst all microlensing surveys to date have been ground based, a survey conducted from space is needed to truly open up the cold exoplanet parameter space. The probability of a detectable planetary signal and its duration both scale as proportional to $\sqrt{\mpl}$, but given the optimum alignment planetary signals from low-mass planets are still quite strong. The lower mass limit for planets to be detectable via microlensing is reached when the planetary Einstein radius becomes smaller than the projected radius of the source star \citep{Bennett:1996emp}. The ${\sim}5.5$-$\mearth$ planet detected by \citep{Beaulieu:2006fem} is near this limit for a giant source star, but most microlensing  events have G or K-dwarf source stars with radii that are at least 10 times smaller than this. In order to extend the sensitivity to Earth mass and below, it is critical to be able to monitor these small source stars that are unresolved from the ground. The ideal machine is a wide field imager in space with sub arsecond imaging capability.

The advantages of undertaking a microlensing exoplanet survey from space (also discussed in Section~\ref{micro}) were first highlighted some time ago by the study of \cite{Bennett:2002ssm} who looked at the potential science from the {\it Survey for Terrestrial ExoPlanets (STEP)} and {\it Galactic Exoplanet Survey Telescope (GEST)}  mission proposals. Building on these proposals, the {\it Microlensing Planet Finder} was proposed to the NASA’s Discovery Program in 2006~\citep{Bennett:2010dwp}. Having realized the synergies between the requirements for cosmic shear measurement and microlensing planet hunting, a microlensing program was proposed as an additional survey as part of the Legacy Science of the {\it Dark UNiverse Explorer} (\dune{}) submitted to ESA Cosmic Vision in 2007 \citep{Refregier:2009, Refregier:2010}.  Ever since, dark energy and microlensing have been advocated for in a joint mission with white papers \citep{Beaulieu:2008pwp} and at international conferences and within the community \citep{Beaulieu:2010eph, Beaulieu:2011}. Our objective is to do a full statistical census of exoplanets down to the mass of Mars from free floating to the habitable zone in complements to the census from the Kepler mission. \dune{} got rebranded into \euclid{} and has been selected as the ESA M2 mission in October 2011, with a statistical census on exoplanets via microlensing being part of the proposed additional survey in the legacy science.

The idea promoted in Europe since 2006 of using a single space telescope to conduct 
both a weak-lensing dark energy survey and an exoplanet
microlensing survey was also followed up in
the US in a number of white papers and conferences
\citep{Bennett:2009, Bennett:2010dwp, Gaudi:2009}. In US the Exoplanet
Task Force report \citep{2008AsBio...8..875L} to the Astronomy and Astrophysics
Advisory committee concluded that ``Space-based microlensing is the
optimal approach to providing a true statistical census of planetary
systems in the Galaxy, over a range of likely semi-major axes''.
Following this the US Astronomy 2010 Decadal Survey endorsed a 
combined approach when it top ranked \wfirst{} \citep{decadalreview,
Barry:2011, wfirstir, Green:2012, Dressler:2012}. The
subsequent report on implementing recommendations of the Decadal Review
\citep{national2012Report} acknowledges that \euclid{} is also capable of undertaking an exoplanet
microlensing survey.

Whilst dark energy studies represent the core science of \euclid{} it also aims to undertake other legacy science. The possibility of an exoplanet survey is mentioned explicitly in the \euclid{} Red Book \citep{redbook} and is currently under study by the \euclid{} Exoplanets Working Group. This paper presents a baseline design for the Exoplanet \euclid{} Legacy Survey (\exels{}). The design for \exels{} is being developed using a detailed microlensing simulator, \mabuls{}, which is also presented in this paper. We focus our attention in the present study exclusively to how \exels{} will probe the cold exoplanet population through microlensing, but \exels{} will also be able to detect hot exoplanets and sub-stellar objects through their transit signatures. This hot exoplanet science is explored separately in a companion paper \citep[][hereafter referred to as Paper~II]{iain}. \exels{} will be the first exoplanet survey designed to probe exoplanets over all host separations, including planets no longer bound to their host. \exels{} will provide an unparalleled homogeneous dataset to study exoplanet demographics and to inform planet formation theories.

We begin the paper by outlining the conservative approach we take to our estimates. In Section~\ref{micro} of this study we overview the basic theory behind exoplanet detection with microlensing and we also describe the \euclid{} mission and its primary science objectives. In Section~\ref{simulator} we introduce our microlensing simulator (\mabuls{}), we describe the \besancon{} population synthesis model Galaxy used to generate artificial microlensing events, and we also outline a baseline design for \exels{}. Section~\ref{yields} presents the results of a simulation of the baseline design for \exels{} and Section~\ref{variations} considers the effects of a number of variations and de-scope options to the baseline design. We end with the summary discussion in Section~\ref{discuss}.

\section{A conservative approach}

Throughout our study of the capability of \euclid{} for detecting exoplanets we adopt a 
conservative approach. There are two reasons for this. Firstly, the design of \euclid{} is itself still evolving. Secondly, since time on a space telescope is expensive, a feasibility study such as carried out in this paper must demonstrate that key science goals are likely to be achieved rather than merely being an aspiration. 

This means that, wherever possible, we aim to make detailed predictions anchored to models which
are known to agree with current data. Where details of models require some assumptions these assumptions must not be overly-optimistic. An example of this approach is our simulation of photometry. The most accurate representation of the photometric methods that will be used on the real data would be to simulate PSF fitting or weighted aperture photometry. However, crowded field photometry is notoriously difficult, and there will always be cases where automated data analysis pipelines will fail to perform the photometry optimally. Simulating all the possible complications in the photometry is impossible. If we were to simulate PSF fitting or weighted aperture photometry, complications that degrade or destroy the photometry would not be modelled and the assumption we had made, while being accurate would be optimistic. Instead, we choose to simulate the photometry as unweighted aperture photometry (see section~\ref{photometry} for full details). Aperture photometry is a less accurate representation of the actual data analysis methods that will be used, and is less effective than optimal photometry by a small but sometimes significant amount. However, this choice is conservative, and helps to ensure that we do not overpredict the performance of the mission.

Given our conservative approach we can have confidence that the scientific yields we predict
are realisable with \euclid{}.

\section{Exoplanetary microlensing from space} \label{micro}

Gravitational microlensing describes the transient deflection and distortion of starlight on milli-arcsecond scales by intervening stars, stellar remnants or planets \citep[for a recent review see][]{Mao:2008iml}. Microlensing is distinguished from ordinary gravitational lensing in that whilst multiple images are produced, they are not resolvable. Instead one observes a single apparent source which appears magnified by a factor
\be 
A = \frac{u^2 + 2 }{u \sqrt{u^2 + 4}},
\ee
where the impact parameter $u$ is a dimensionless angular separation between the lens and source measured in units of the angular Einstein radius of the lens.
A microlensing event is observable as a transient achromatic brightening of a background source star lasting for $\tein \sim 6-60$~days, where $\tein$ is the Einstein radius crossing time. For a single lens the lightcurve profile is time-symmetric, with a peak magnification occurring when the impact parameter is at its minimum $u = u_0$. The lensing signal from foreground stars is detectable in a few out of every million background stars located in crowded stellar fields such as the Galactic bulge. A planet orbiting a foreground lensing star may, in a few percent of microlensing events, perturb the microlensing signal causing a  brief deviation which lasts for $t_{\rm p} \sim \tein \sqrt{M_{\rm p}/M_*}$, where $M_*$ and $M_{\rm p}$ are the host and planet masses. Typically $t_{\rm p}$ is in the region of a day for a Jupiter mass planet down to a few hours for Earth mass planets. The intrinsically very low probability $\sim {\cal O}(10^{-8})$ of an exoplanetary microlensing signature against a random background source star, coupled with the brief deviation timescale associated with Earth-mass  planets, places huge technical demands on microlensing surveys. 

The probability
of a planetary perturbation occurring scales roughly as the square root of the
planet mass, or more strictly, as the square root of the planet-host
mass ratio $q$~\citep{Gould:1992pmm}. This shallow sensitivity curve
makes microlensing ideal for detecting low-mass planets. The scaling
breaks down below about a Mars mass, where finite-source effects begin
to wash-out planetary signatures, even for main-sequence source stars~\citep{Bennett:2002ssm}.

The
sensitivity of microlensing to planets peaks close to the Einstein radius $\re$ with projected
semimajor axis $a_{\perp}\sim\re\sim 2$~AU, corresponding to where the microlensing images are
most likely to be perturbed~\citep{Wambsganss:1997pmm,
  Griest:1998cwd}. However there is significant sensitivity to planetary orbits with
$a_{\perp}\sim 0.5$~AU, and outwards to infinity \citep[i.e.
free-floating planets][]{Han:2004ffp,Sumi:2011ffp}.

Owing to its high stellar density
and microlensing optical depth, the Galactic bulge is the best target for microlensing studies. Towards the
bulge, extinction is a significant problem at optical wavelengths. Additionally, the extreme stellar crowding and
arcsecond-scale seeing mean that only the giant star population can
be properly resolved from the ground \citep{Bennett:2004gsm}. Observing in the near-infrared lessens the effects of dust and so provides a larger microlensing optical depth \citep{Kerins:2009smm} but, from the ground, stellar crowding problems are even more severe, and noise levels are enhanced due both to the sky and unresolved stellar backgrounds. Therefore, in order to
monitor enough source stars, ground-based surveys must regularly cover ${\sim}100$~deg$^{2}$. Current and future ground-based
surveys -- e.g., MOA-II~\citep{Sumi:2010mii},
OGLE-IV~\citep{Udalski:2011og4}, KMTNet~\citep{Kim:2010kmt} and
AST3~\citep{Yuan:2010ast3} -- with wide-field imagers will achieve
suitable cadence and areal coverage to detect routinely large
numbers of giant planets if they exist in sufficient abundance.
However they will not be able to monitor enough stars at high-cadence
to detect Earth-mass planets at a significant rate. For this
reason, targeted follow-up of promising microlensing events by large
networks of small telescopes is currently used to achieve high cadence and
continuous event coverage~\citep[see, e.g.,][]{Gould:2010pps}, and
to push the sensitivity of ground-based microlensing firmly into the
super-Earth regime~\citep{Beaulieu:2006fem,Bennett:2008lmp}. However,
the follow-up networks only have the capacity to observe ${\sim} 100$
events per year or fewer with sufficient cadence \citep{Peale:2003gsm}. This allows the mass
function to be probed down to ${\sim} 5$--$10\mearth$, and possibly the semi-major
axis distribution of planets above ${\sim} 50\mearth$, but is unlikely
to provide more than isolated detections below these masses~\citep{Peale:2003gsm, Bennett:2004gsm,
Dominik:2011pmf}.

Observations from space are able to overcome many of the problems
facing ground-based observers. A space telescope has better resolution
due to the lack of atmosphere and also a lower sky background,
especially in the infrared. This means that with appropriate
instrumentation, a space telescope can resolve main-sequence sources
in the bulge and monitor the required ${\sim} 10^8$ sources over a much
smaller area. This in turn allows high-cadence observations on a small
number of fields~\citep{Bennett:2002ssm, Bennett:2004gsm}. The
fundamental requirements of a space telescope for a microlensing
survey are a wide field of view ($\gtrsim 0.5$~deg$^2$), with a small
pixel scale. In order to minimize the effect of extinction towards the
Galactic bulge, it should observe in the near infrared. The telescope
must also have a large enough collecting area to allow high-precision
photometry of main-sequence bulge stars in short exposure times. These
are almost exactly the same requirements as for dark energy studies using weak lensing, which are already driving the hardware design of \euclid{}.

\subsection{The \euclid{} mission}

\euclid{} is an M-class mission within the ESA Cosmic Vision programme. It aims to investigate the nature of
dark energy through
measurements of weak gravitational lensing and baryon acoustic
oscillations \citep{redbook}. \euclid{} will comprise a $1.2$-m Korsch telescope with a high-resolution optical imager (\vis{}) and a near infrared imaging
spectrometer (\nisp{}), operating simultaneously. The core science mission will involve a
$15000$-deg$^2$ wide survey and $40$-deg$^2$ deep survey over six
years to measure galaxy shapes and photometric and spectroscopic
redshifts. \vis{} will observe with a wide optical band-pass covering
$R$, $I$ and $Z$, and \nisp{} will have available three infrared filters: $Y$,
$J$ and $H$. The currently envisaged step and stare survey strategy of \euclid{} means
that for up to two months per year it will point towards the Galactic plane and away from its primary 
science fields. As stated in \cite{redbook} it is intended that some of this time will be devoted to other legacy science. A planetary microlensing survey is one option described in \cite{redbook} and is being actively evaluated by the \euclid{} Exoplanets Working Group.

The similarity of hardware requirements for dark energy and exoplanet microlensing space missions has been recognised for some time \citep{Bennett:2002ssm}, and most recently by the 2010 US Astrophysics Decadal Review \citep{decadalreview}. This review recommended the merger of three mission concepts into one mission, the Wide-Field Infrared Survey Telescope \citep[\wfirst{}, ][]{Green:2012}. Two of the core science objectives for \wfirst{} are a dark energy survey and an exoplanets survey using microlensing. In the baseline \wfirst{} concept the microlensing survey will total $432$ days, somewhat longer than will be feasible for \exels{}.

\section[\mabuls{}]{The Manchester-Besan{\c c}on microLensing Simulator (\mabuls{})}\label{simulator}

We have designed the Manchester-\besancon{} micro-Lensing Simulator
(\mabuls{} -- pronounced {\it may}-buls) to perform detailed simulations of the \exels{} concept. 
\mabuls{} is the first microlensing simulator to use a combination of
a population synthesis Galactic model with a realistic treatment of
imaging photometry. This means that every aspect of the simulation,
including the event rate calculations, blending and photometry are
simulated self-consistently.

Several key ingredients are needed in order to simulate any microlensing survey. A simulator
must draw its simulated events from a Galactic model and distributions
of the event parameters. It must simulate the observations of the
survey, and finally, it must also simulate the detection criteria used
to select its sample of events. It is also necessary to make a
choice as to the complexity of the microlensing model used to simulate
events. For example, is the lens composed of a single mass or multiple
components? Are higher-order effects such as parallax and orbital
motion included? In the rest of this section we will discuss both how \mabuls{}
implements each component of the simulation and the choice of
parameters we use in the simulation of \exels{}. Unless stated otherwise, we have taken the survey parameters
from the \euclid{} Red Book \citep{redbook}. 

\subsection{The \besancon{} Galactic model} \label{galsim}

Underpinning the \mabuls{} microlensing event generation is the \besancon{} model~\citep{Robin:1986bgm, Robin:2003bgm, Robin:2012bgm}, a population synthesis model of the Galaxy. 
The \besancon{} model comprises five main stellar populations, a
spheroid (stellar halo), thin and thick discs, a bar and bulge. The
stars of each population are assumed to be formed from gas for input models of
star formation history and initial mass function (IMF). The stars are aged according to 
model evolutionary tracks to their present-day
state~\citep{Haywood:1997set}. This determines the
distribution of stellar bolometric fluxes, which are converted to colours and
magnitudes using stellar atmosphere models convolved with standard band-pass templates in 
various photometric systems. 

The spatio-kinematic
distribution of the disc stars is determined by integration of a
self-consistent gravitational model using the Poisson and Boltzmann
equations. Finally, the observed colours and magnitudes are corrected for extinction
using a three-dimensional dust model~\citep{Marshall:2006ged}. A
limited number of model parameters are then
optimized to reproduce observed star counts and kinematics. The output
of the model is an artificial catalogue of stellar photometry and kinematics for a survey of 
specified sensitivity and areal coverage.

The \besancon{} model is in constant
development~\citep[e.g.,][]{Robin:2012bgm}. In this work we
use version 1106 of the \besancon{} model, though an updated version of
the model has been released since. In subsequent models, the
properties of the Galactic bar and bulge (see below) change somewhat from those we
use here. Below we briefly overview the properties of the main stellar components used to generate
microlensing events in \mabuls{}. The Solar Galacto-centric distance in the model is 8~kpc.

\subsubsection{The stellar halo}

The stellar halo is modelled as being formed by a single burst of star
formation at $14$~Gyr, with metallicity centred at $[{\rm Fe}/{\rm H}]=-1.78$ and with a dispersion of 0.5. It
has a triaxial velocity distribution with dispersions $(\sigma_U,
\sigma_V, \sigma_W) = (131, 106, 85)$~\kms. Its density is small everywhere, even at the Galactic center, and so contributes only marginally to the optical
depth and microlensing event rate. 

\subsubsection{The bar}

The bar, altered from the bulge-like component used by \citet{Kerins:2009smm}, consists
of a boxy triaxial distribution, similar to that described
by~\citet{Picaud:2004bmb}, but with a Gaussian density law as opposed
to a \citet{Freudenreich:1998gbm} $\operatorname{sech}^2$ law~\citep{Robin:2012bgm}. The major axis of the
triaxial structure lies at an angle of $12.5\degr$ relative to the
Sun--Galactic centre line of sight and has scale lengths
$(X,Y,Z)=(1.63,0.51,0.39)$~kpc, where the $X$ direction is parallel to
the major axis and the $X$ and $Y$ axes lie in the Galactic
plane. It is truncated at a Galactocentric radius of
$2.67$~kpc. The bar rotates as a solid body with a speed of
$40$~\kms~kpc$^{-1}$. The velocity dispersions in the bar along the
axes defined above are $(113,115,100)$~\kms. The central stellar mass
density of the bar, excluding the central black hole and clusters, is $19.6 \times 10^{9}\msun$~kpc$^{-3}$. 

Embedded within the bar is also an additional component~\citep[somewhat different from the ``thick bulge'' component in][]{Robin:2012bgm}. However, in the version of the model we
use here, its density is smaller by ${\sim} 10^{-4}$ times that of the bar, so we do not describe it further. We use the terms ``bar'' and ``bulge'' inter-changeably from here onwards to mean the bar component of the \besancon{} model.

The stellar population of the bar is assumed to form in a single
burst $7.9$~Gyr ago~\citep{Picaud:2004bmb},
following~\citet{Girardi:2002tic}. The bar IMF ($\dd N/\dd M$)
scales as $M^{-1}$ between $0.15\msun$ and $0.7\msun$, and follows a Salpeter slope above
this mass. The population has a mean metallicity [Fe/H]$=0.0$ with
dispersion $0.2$ and no metallicity gradient. The stellar
luminosities are calculated using Padova
isochrones~\citep{Girardi:2002tic}.

\subsubsection{The thick disc}

The thick disc is modelled by a single burst of star formation at
$11$~Gyr. Its properties have been constrained using star counts by
\citet{Reyle:2001tdm}. The thick disc contributes only marginally to
the microlensing event rate, so we do not describe it in detail. Its
parameters are described by \citet{Robin:2003bgm}.

\subsubsection{The thin disc}

The thin disc is assumed to have an age of $10$~Gyr, over which star
formation occurs at a constant rate. Stars are formed with a two-slope
IMF that scales as a power-law $M^{-1.6}$ from $0.079\msun$ to $1\msun$ and $M^{-3}$ above,
based on the {\it Hipparcos} luminosity
function~\citep[e.g.,][]{Haywood:1997set}, with updates described by~\citet{Robin:2003bgm}. Stars below $1\msun$ follow the evolutionary
tracks of \citet{Vandenberg:2006set}, while those above follow
\citet{Schaller:1992sem} tracks. The thin disc follows an
\citet{Einasto:1979gmm} density profile with a central hole. The
density normalization, kinematics and metallicity distribution of the
disc depend on stellar age, with seven age ranges defined, whose
parameters are given by \citet{Robin:2003bgm}. The Solar velocity is
$(U_{\sun},V_{\sun},W_{\sun})=(10.3,6.3,5.9)$~\kms, with respect to the local
standard of rest $V_{\mathrm{LSR}}=226$~\kms. The disc has a scale length $2.36$~kpc,
and the hole has a scale length $1.31$~kpc, except for the youngest
disc component which has disc and hole scale lengths of $5$~kpc and
$3$~kpc, respectively.  The disc is truncated at
$14.0$~kpc. The scale height of the disc is computed self-consistently
using the Galactic potential via the Boltzmann equation as described
by \citet{Bienayme:1987mdg}. Also modelled in the disc are its warp
and flare~\citep{Reyle:2009bmd}.

\subsubsection{Extinction}

Extinction is computed using a three-dimensional dust distribution
model of the inner Galaxy ($|\ell|<100\degr$, $|b|<10\degr$), built by
\citet{Marshall:2006ged} from analysis of 2MASS
data~\citep{Cutri:2003asc} using the \besancon{}
model. \citet{Marshall:2006ged} did this by comparing observed,
reddened stars to unreddened simulated stars drawn from the \besancon{}
model. From this the extinction as a function of distance along a given
line of sight is computed by minimizing $\chi^2$ between observed and
simulated $J-K_{\mathrm{s}}$ colour distributions. The resulting map
has a ${\sim} 15$-arcmin resolution in $\ell$ and $b$, and a distance
resolution ${\sim} 0.1$--$0.5$~kpc, resulting from a compromise between
angular and distance resolution.

\subsubsection{Other components}

The \besancon{} model also takes account of other Galactic components,
including the mass due to the dark matter halo and interstellar
medium. The details of these components are given by
\citet{Robin:2003bgm}. White dwarfs are included in the model
separately to normal stars, with separate densities and luminosity
functions determined from observational constraints~\citep[][and
  references therein]{Robin:2003bgm}. The evolutionary tracks and
atmosphere models of \citet{Bergernon:1995wdm} and
\citet{Chabrier:1999wdm} are used to compute their colours and magnitudes.

\subsection{Microlensing with the \besancon{} model}

Following the method of \citet{Kerins:2009smm}, \mabuls{} uses two star
lists output by the \besancon{} simulation to construct catalogues of
possible microlensing events and calculate their properties. The first
list, the source list, is drawn from the \besancon{} model using a
single magnitude cut in the primary observing band of the survey. A
second list, the lens list, is drawn from the model without a magnitude
cut. Both source and lens lists are truncated at a distance of
$15$~kpc to improve the statistics of nearer lenses and sources that
are much more likely to be lensed/lensing.

\begin{table}
\caption{The magnitude range, and the average number of stars $\langle N_{\star} \rangle$ in the \besancon{} model star catalogues used in this work. Bright, moderate and faint are the three levels of catalogues used to build the simulated images.}
\begin{tabular}{@{}lccc}
Catalogue & Mag. range & Solid angle (deg$^2$) & $\langle N_{\star} \rangle$ \\
\hline
Source   & $10<H_{\rm vega}\le 24$ & $2\times 10^{-4}$ & 23219 \\
Lens     & $-\infty<H_{\rm vega}\le \infty$ & $2\times 10^{-4}$ & 32933 \\
Bright    & $-\infty <H_{\rm vega}\le 15$ & $1\times 10^{-3}$ & 441 \\
Moderate & $15 < H_{\rm vega} \le 24$ & $2\times 10^{-5}$ & 2312 \\
Faint   & $24 < H_{\rm vega} \le \infty$ & $2\times 10^{-5}$ & 967 \\
\end{tabular}
\label{catsize}
\end{table}

Overall microlensing event rates are calculated along multiple
lines of sight, with spacing set by the resolution of the
\citet{Marshall:2006ged} dust map. The total rate due to each pair of
source and lens lists, about the line-of-sight $(\ell, b)$, is
\be
\Gamma(\ell,b) =
\frac{\Omega_{\mathrm{los}}}{\delta\Omega_{\mathrm{s}}}\sum^{\mathrm{Sources}}
  \left(\frac{1}{\delta\Omega_{\mathrm{l}}}\sum_{\dl<\ds}^{\mathrm{Lenses}}
  2\thetae\murel\right),
\label{rate}
\ee
where $\Omega_{\mathrm{los}}$ is the solid angle covered by a dust-map
resolution-element, and $\delta\Omega_{\mathrm{s}}$ and
$\delta\Omega_{\mathrm{l}}$ are the solid angles over which the source
and lens catalogues are selected, respectively. The rate is calculated
over all possible source-lens pairs to minimize the noise of
counting statistics. The average sizes of all the catalogues used in this work are listed in Table~\ref{catsize}.

To simulate microlensing, \mabuls{} draws sources and lenses from their
respective lists with replacement, requiring the source be
more distant than the lens. From the source and lens parameters, the
Einstein radius and timescale are computed, as well as the rate
weighting assigned to the event
\be
\gamma = \umax\thetae\murel,
\label{weighting}
\ee
where $\umax$ is the maximum impact parameter of the event; how $\umax$ is
determined is discussed in the following sections. Events are
simulated and those that pass the detection criteria are flagged. The
rate of detections in a given dust-map element is the sum of the weights
of detected events normalized to the sum of the rate weightings for
all the simulated events -- this is essentially a detection
efficiency. The detection efficiency is then multiplied by the total
line-of-sight rate computed in Equation~\ref{rate} to yield the
expected detection rate for $0.25\times 0.25$~deg$^2$, the size of
the dust-map element. These rates are then summed over all the
dust-map elements to yield the total simulation event rate.

For the \exels{} simulations we restrict the source magnitude to a vega-based $H$-band magnitude $H_{\mathrm{vega}}<24$. This corresponds to an AB magnitude limit of $H_{\mathrm{AB}}<25.37$. Unless otherwise noted, AB magnitudes will be used throughout the paper.

\subsubsection{Normalization of the event rate}

The absolute number of stars in the \besancon{} model has been set by normalizing their number density to match star counts along a number of lines of sight that sample the different components of the Galaxy~\citep{Robin:2003bgm}. This process will only be accurate down to the limiting magnitude of the star count data, and it is possible that the number of fainter stars predicted by the model could be incorrect. While these fainter stars may not contribute to star counts, they do contribute to the microlensing event rate, either as lenses or sources. It is therefore important to make sure that the microlensing event rates predicted by the \besancon{} model match those that are observed.

The microlensing event rate is a potentially powerful but complex probe of Galactic structure because it depends on the kinematics of the lens and source populations as well as the lens mass function. Often, surveys aim instead to measure the microlensing optical depth, which is much more cleanly defined as it only depends only on the distribution of lenses and sources along the line of sight. The total event rate $\Gamma$ is related to the optical depth $\tau$ by \citep[e.g., ][]{Paczynski:1996mlr}
\begin{equation}
\Gamma \propto \frac{N_{\star}\tau}{\langle \tein \rangle}
\label{ratetau}
\end{equation}
where $N_{\star}$ is the number of monitored sources and $\langle \tein \rangle$ is the average event timescale. We can verify the microlensing event rate predicted by the \besancon{} model by comparing each of the quantities on the right hand side of Equation~\ref{ratetau} to measured values, and make a correction to the rate if required.

\begin{figure}
\includegraphics[width=\figuresize]{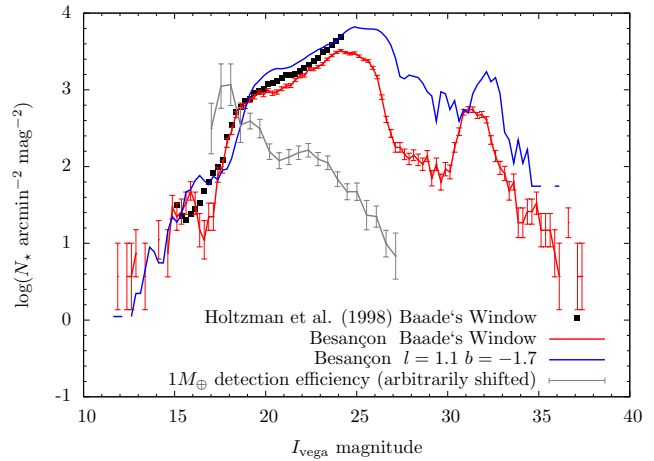}
\caption{Comparison of the luminosity function in the bulge as measured by \citet{Holtzman:1998blf} in Baade's window ($\ell=1$, $b=-3.9$) to that of the \besancon{} model at the same location. The measured luminosity function, shown by the black line, has been returned to the $I$-band apparent magnitude scale by adopting the distance modulus ($\mu=14.52$) and extinction ($A_I=0.76$) values of \citet{Holtzman:1998blf}. The red line shows the \besancon{} model's luminosity function at Baade's window, while the blue line shows the \besancon{} model luminosity function at the center of the ExELS fields ($\ell=1.1$, $b=-1.7$). The grey line shows the Earth-mass planet detection efficiency for Euclid as a function of source $I_{\mathrm{vega}}$ magnitude (abitrarily shifted on the log scale).}
\label{lumfunc}
\end{figure}

In the bulge region, the \besancon{} model has been normalized to star counts from the 2MASS survey~\citep{Cutri:2003asc, Robin:2012bgm}, which is relatively shallow compared to the sources that \euclid{} will observe. The number counts of fainter sources are extrapolated using the IMF and extinction maps, and any uncertainties in these will propogate to the source counts. There is a relative paucity of deep star count measurements in the bulge with which to test the \besancon{} model assumptions, with published measurements only along a single line of sight close to our proposed \euclid{} fields. To assess the validity of $N_{\star}$, we compare this measurement of the luminosity function in Baade's window~\citep{Holtzman:1998blf} to the \besancon{} star catalogue produced for the same line of sight. Figure~\ref{lumfunc} shows both luminosity functions. There is good agreement between the two luminosity functions in the magnitude range $17<I_{\rm vega}<20$, but fainter than this the \besancon{} model under-predicts the number of stars. Integrated over the whole range of the \citet{Holtzman:1998blf} luminosity function, the \besancon{} model predicts $32.46\times 10^6$ stars per square degree, but \citet{Holtzman:1998blf} measure $42.06\times 10^6$ stars per square degree. To correct the event rate for this lack of stars requires multiplying by a factor of $1.30$. While we adopt this correction, we caution that a significant fraction of the discrepancy arises from the faintest part of the luminosity function, where \citet{Holtzman:1998blf} argue that their completeness corrections are uncertain. Beyond the faintest bin of the \citet{Holtzman:1998blf} luminosity function we might worry that the number of stars keeps on increasing in reality, while in the \besancon{} model it begins to fall off, suggesting a larger correction factor would be required. However, average extinction in the \exels{} fields \citep[$A_I=1.73$ at a distance of $8$~kpc][]{Marshall:2006ged} is nearly one magnitude more than $A_I=0.76$, the value adopted by \citet{Holtzman:1998blf} in their Baade's window field. This implies that in the \exels{} fields the \citet{Holtzman:1998blf} luminosity function extends down to $I_{\rm vega}\approx 25.2$. At this source magnitude, the planet detection efficiency for \euclid{} has fallen to roughly a third of the average at brighter magnitudes, and falls rapidly as the sources get fainter. Therefore, while there may be more faint stars that the \besancon{} model is missing, including these source stars will not significantly increase the number of planet detections.

\begin{figure}
\includegraphics[width=\figuresize]{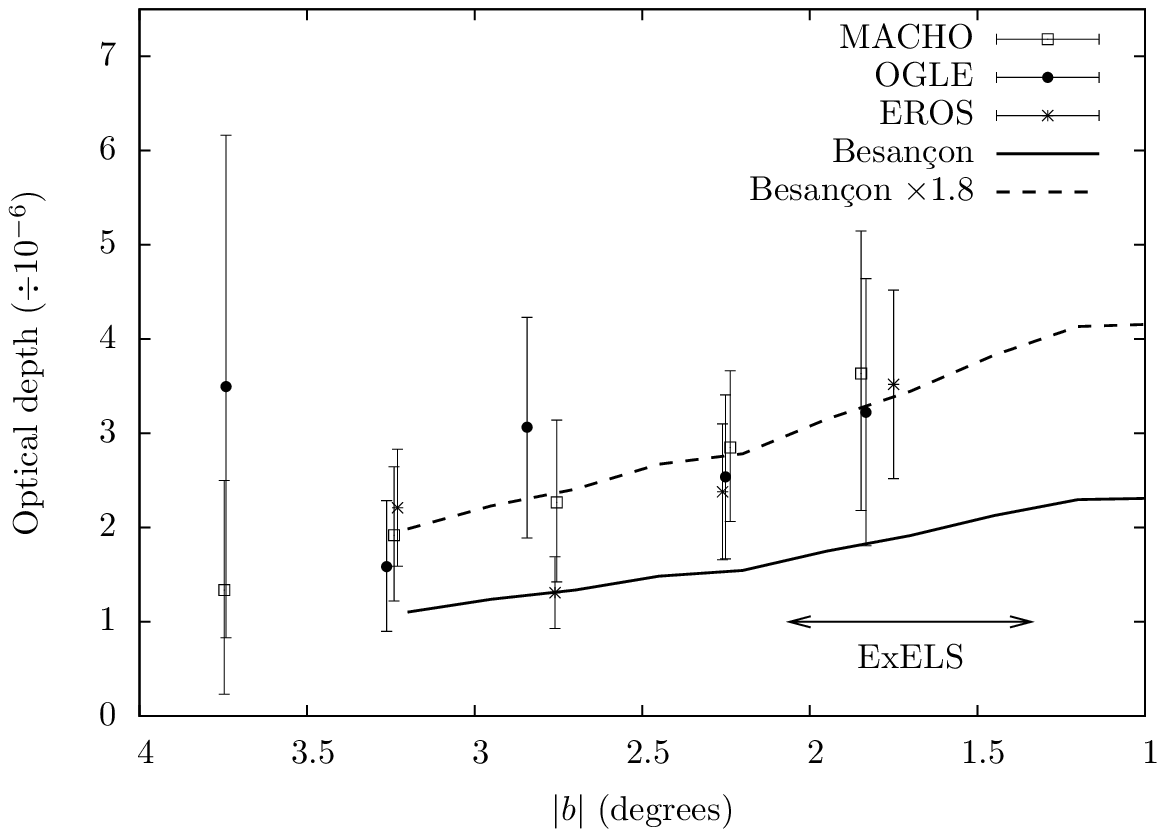}
\caption{A comparison of measured optical depths to red clump giants to those calculated from the \besancon{} model. Open square, filled circle and asterix data points show results from the MACHO~\citep{Popowski:2005mod}, OGLE~\citep{Sumi:2006odm} and EROS~\citep{Hamadache:2006eod} surveys, respectively. The solid line shows the average optical depth to red clump stars selected from the \besancon{} model based on their absolute magnitudes and colours. The dashed line shows the same line as the solid curve, but multiplied by a constant $1.8$ to match the data.}
\label{optdepth}
\end{figure}

The optical depth in the Galactic bulge has been measured to two different source populations: red clump giants and difference imaging analysis (DIA) sources. Measurements of DIA optical depths are typically slightly higher than those for clump giant sources. Clump giants are abundant, bright standard candles, and those in the bulge can be easily recognised and isolated by their position on a colour-magnitude diagram. They therefore make an ideal tracer population of the bulge. DIA sources however, are less clearly defined, as they depend on the sensitivity of the survey. Due to this difficulty of defining the DIA source population, we only compare the \besancon{} model to measured red clump optical depths, which have been measured by the MACHO~\citep{Popowski:2005mod}, EROS~\citep{Hamadache:2006eod} and OGLE~\citep{Sumi:2006odm} surveys. Figure~\ref{optdepth} shows the red clump optical depth measurements of each of these surveys as a function of absolute Galactic latitude, together with the average optical depth to red clump stars as predicted by the \besancon{} model. The \besancon{} optical depths are averaged over the longitude range $-0.525<\ell<2.725$. It is clear that the \besancon{} optical depth is somewhat lower than the measured optical depth. Multiplying the \besancon{} model optical depths by a constant factor of $1.8$, brings the predictions into good agreement with the measurements.

The average timescale reported by microlensing experiments is somewhat ill-defined due to the arbitrary timescale cut-offs applied to their event samples, but are typically in the range of $20$--$30$~days. The average timescale calculated from the timescale distribution presented in Figure~\ref{timescales} is $21.4$~days without applying any detection criteria and $29.2$~days for events detected above baseline. The average timescale of the \citet{Sumi:2011ffp} sample is $26.0$~days. The sample detected above baseline is most comparable to the \citep{Sumi:2011ffp} sample, but is not directly comparable. As the difference is of the order of $10$--$20$~percent, and the samples are not directly comparable, we choose not to make a correction to the event rate based on the timescales.

Combining the required correction factors for optical depth and source counts, we conclude that the microlensing event rates predicted by the \besancon{} model likely require a correction factor
\begin{equation}
f_{1106} = 1.8\times 1.30 = 2.33,
\label{corrfac}
\end{equation}
where the subscript emphasizes the fact that this scaling is only applicable to the version of the \besancon{} model we are using. We have multiplied all raw results by this factor throughout the paper. To account for further uncertainty in the overall event rate we define an additional multiplicative factor $\rtau$, with which all the event rates and numbers of planet detections should be multiplied. In this paper we advocate for a value $\rtau=1$. Note, however, that \citep{Green:2012} choose a value $\rtau=1.475$ to compromise between different values of the planet detection efficiency determined from \mabuls{} and simulations following \citet{Bennett:2002ssm}.

One possible cause of the low predicted optical depth could be missing low-mass stars and sub-stellar objects too faint to be included in star counts. The lower cut-off of the bulge mass function in the \besancon{} model is $0.15\msun$. Extending the mass function down to ${\sim}0.03\msun$, keeping the same low-mass slope ($M^{-1}$), would account for the missing optical depth \citep[see][for a discussion of the effect of the mass function of microlensing event rates]{CalchiNovati:2008}. Adding such low-mass stars to the star catalogues would increase the number of planet detections by a factor larger than the increase in optical depth, because the mass ratio of planets around these stars would be larger than the mass ratio of planets around a star of the average stellar mass in the catalogues. Another possible cause of the low optical depth is the lack of high-mass stellar remnants -- neutron stars and black holes, which are not included in the \besancon{} catalogues. Should these be the cause of the low optical depth, the number of planet detections would not increase as much as the optical depth, because even if planets remained around these objects, their mass ratios would be smaller than the average in the unmodified catalogues. Other possible causes of the optical depth discrepancy, such as problems with Galactic structure, would cause the number of planet detections to change in the same way as the microlensing event rate. Note that in all the scenarios discussed here, the average timescale is affected -- adding low-mass stars decreases the average timescale, giving a further boost to the event rate, while adding high-mass remnants increases the average timescale, further supressing any boost. Comparing the \besancon{} timescale distribution to that observed by \citet{Sumi:2011ffp} (see Figure~\ref{timescales}) suggests that the \besancon{} model is missing short timescale events, but each timescale distribution has different selection criteria, so it is impossible to draw firm conclusions.

\subsection{The microlensing events} \label{planetsim}

\mabuls{} uses user-supplied functions to compute microlensing
lightcurves including any effects that the user wants to model. For
this work we modelled only planetary lens systems composed of a
single planet orbiting a single host star. As we want to investigate
the planet detection capability of \exels{} as a function of planet mass
$\mpl$ and semimajor axis $a$, we chose to simulate systems with
various fixed values of planetary mass in the range $0.03 - 10^4~\mearth$ and semimajor axis distributed
logarithmically in the range $0.03 < a < 30$~AU. We assume a
circular planetary orbit that is inclined randomly to the line of
sight. The orbital phase at the time of the event is again random; we do 
not model the effects of orbital motion in the
lens. The impact parameter and angle of the source trajectory are
distributed randomly, with the impact parameter in the range 
$\uzero = 0$--$\umax$. For \exels{} simulations we choose $\umax=3$. While it may be the case that some of the stellar microlensing events with $\uzero\approx 3$ will not be detected, it is still possible for planets to cause detectable signals.

The planetary microlensing lightcurves are computed assuming that the
source has a uniform intensity profile (in other words, no limb
darkening). Test simulations including the effect of limb darkening (which is small in the infrared) showed that inclusion of the effect would change the number of Earth-mass planet detections by less than $1$~percent. The finite-source magnification is computed using the
hexadecapole approximation when finite-source effects are
small~\citep{Pejcha:2009fsa, Gould:2008hfs} and the contouring
method when they are not~\citep{Gould:1997fss,
  Dominik:1998fsc}. Finite-source effects are accounted for in
single-lens lightcurve calculations using the method
of~\citet{Witt:1994fs}. When fitting lightcurves with the single-lens
model, we use a finite-source single-lens model if the impact
parameter $\uzero < 10 \rhostar$, where $\rhostar$ is the ratio of
angular source radius to the angular Einstein radius. Otherwise the
point-source single-lens model is used.

\subsubsection{Free-floating planets}

Observations from nearby star clusters, as well as tentative evidence from ground-based microlensing surveys, suggests that planets can occur unbound from any host, sometimes referred to as free-floating planets. If free-floating planets exist in significant numbers then \exels{} should detect them as relatively brief single-lens microlensing events.

At this stage we have no clear information to allow us to characterise a Galactic population of free-floating planets with confidence. What we know from young star clusters like sigma Orionis is that brown dwarfs ($0.072-0.013 ~\msun$) and massive free-floating planets ($0.013-0.004~\msun$) are as numerous a low-mass stars with masses in the interval $0.25-0.072~\msun$ \citep{PenaRamirez:2012}. However, given the uncertainties over the characteristics of a Galactic-wide distribution of planets we adopt a simple scalable assumption of one free-floating planet of mass $\mffp$ per Galactic star. 

As free-floating planets are single, point-mass lenses we treat them
in a separate \mabuls{} simulation. Each lens star drawn from the
\besancon{} simulation is replaced by a planet of mass $\mffp$. We simulate a range of values for $\mffp$ from $0.03 - 10^4 ~\mearth$. We assume
the planets retain the same kinematics, but the fundamental microlensing
properties such as the Einstein radius change to reflect the reduced
mass. Ejected planets may well have a somewhat larger velocity dispersion than their original hosts, in which case the rate of free-floating planet events increases proportionately and the timescale decreases inversely with their velocity. We assume that free-floating planets emits no detectable light, which is a good assumption for typical distances at which planets are detectable through microlensing. Each
lightcurve is calculated using the finite-source single-lens
model. The impact parameter is chosen to lie in the range
$\uzero=0$--$\umax$, where we choose $\umax=1$ to remain conservative, and 
we require that the time
of peak magnification lie within an observing season (unlike for the
standard simulations).

\subsection{\euclid{} observing strategy} \label{obsstrat}

The \exels{} survey must be capable of detecting planets at least down to Earth masses, which means we require
an observing cadence of no more than 20 mins between repeat observations of the same field. It must also monitor enough
source stars over a sufficiently long observing baseline to ensure a healthy detection rate. As shown by \cite{Kerins:2009smm} the event rate is optimised at near-infrared wavelengths, suggesting that the \nisp{} camera should be the primary instrument for \exels{}. In Section~\ref{bands} we show that this is indeed the case, despite the significantly worse resolution of the \nisp{} instrument relevant to the optical \vis{} instrument.

In order to achieve a cadence of no worse than 20~min, \exels{} will be able to monitor up to 
3 target fields of ${\sim} 0.5$~deg$^2$ with a total exposure of $270$~s per
pointing, split into stacks of $3$ ($Y$- and $J$-band) or $5$
($H$-band) exposures with \nisp{}. We assume that there
is $5$~s of dead time between the exposures of a stack. The \vis{}
instrument pointings consist of a single $540$-s exposure. 
We assume a baseline slew and settle time of $85$~s, though in Section~\ref{slewtime} we also consider the effect of a substantially longer slew and settle time.
We assume that any readout, filter wheel rotation and data down-link is
performed during slewing. Some of these parameters are summarized in
Table~\ref{parameters}. 

\begin{figure}
\includegraphics[width=\figuresize]{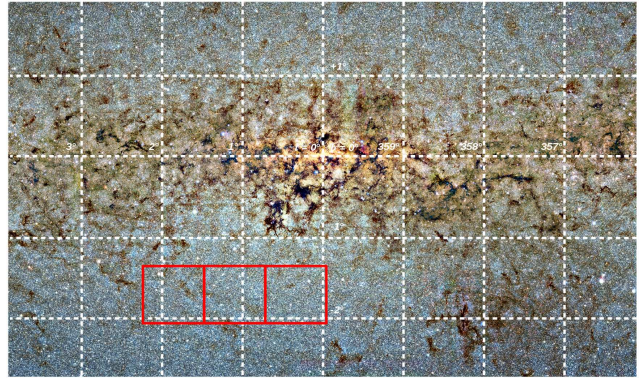}
\caption{The approximate location of the three \exels{} field pointings (solid line rectangles) assumed for the simulation, with the middle of the three fields centred at $l = 1.1^{\circ}$, $b = -1.7^{\circ}$. Each \exels{} field has dimensions of $0.76 \times 0.72$~deg$^2$. The background image of the Galactic Centre is a near-infrared mosaic of images from the VVV survey \protect{\citep{Saito:2012vvv}}. Background image credit: Mike Read (Wide-field Astronomy Unit, Edinburgh) and the VVV team.}
\label{fields}
\end{figure}

We simulate a total observing baseline of
$300$~days for \exels{}, spread over $5$~years with two $30$-day seasons per year. This strategy is determined by the design of the spacecraft's sun shield. This restricts \euclid{} to observing fields with solar aspect angles between $89$ and $120$ degrees. As the Galactic bulge lies near to the ecliptic, \euclid{} can only observe bulge fields uninterrupted for up to $30$ days, twice per year. It should be stressed that a 10-month survey represents a firm theoretical maximum that could be possibly achieved over 5 years. In practice the cosmology primary science will likely prohibit much legacy science being undertaken in the first few years of the mission, so that a 6-month exoplanet survey likely represents a more achievable goal during the 6-year primary cosmology mission. It is possible that, if \euclid{} remains in good health beyond 6 years, a full 10 month programme could be completed after the cosmology programme is complete. We therefore investigate the exoplanet science returns possible for a survey of up to 10-months total time. We assess the impact of shorter total baselines in Section~\ref{yields}.

For the \exels{} simulation we use three contiguous \euclid{} pointings aligned parallel to the Milky Way plane, with the central field located at Galactic coordinates $l = 1.1^{\circ}$, $b = -1.7^{\circ}$ (J2000), as shown in Figure~\ref{fields}. Each \euclid{} field covers $0.76 \times 0.72$~deg$^2$ on the sky, giving a total \exels{} survey area of 1.64~deg$^2$. We conservatively assume most of the observations
are taken with \nisp{} in only one filter (we show in Section~\ref{variations} that $H$-band is the best filter choice), at a cadence of roughly one observation every 20 mins. Conservatively we add colour information from the two other \nisp{} filters and the \vis{} camera at a rate of only one observation every 12 hours. This conservative assumption guarantees we will not be limited by telemetry rate restrictions. However, in Section~\ref{bands} we consider the benefits of simultaneous \nisp{} and \vis{} imaging. For the hot exoplanet science investigated in Paper~II we note that it is important to achieve high cadence observations with both \vis{} and \nisp{} instruments. Therefore strong limitations in telemetry could impact somewhat upon the hot exoplanet science but is unlikely to impact significantly upon the cold exoplanet science discussed here.

The number of planets which can be detected by \exels{} will be governed by the overall rate of microlensing within the survey area, though only a small fraction of these will have detectable planetary signatures.
The expected overall number of microlensing events (with or without planet signatures) that would be detected as significantly magnified by \exels{} is ${\sim}27000$~events with $\uzero\le 3$ over the course of a 300-day survey. This is in excess of the total number of microlensing events discovered by all ground-based microlensing surveys since they started operations twenty years ago. However some of these events 
may not be well characterised if they peak outside of one of the observing seasons. Placing the restriction that the time of peak magnification, $\tzero$, must be contained within one of the observing seasons lowers the overall number to ${\sim} 9800$~events (${\sim} 5700$~events with $\uzero<1$) for a 300-day campaign, or about 1000~events per month. This is an improvement of a factor of ${\sim 65}$ in detection rate per unit time per unit area over the OGLE-IV survey \citep{Udalski:2011og4} in its best field, which yields ${\sim 15}$ events per month per deg$^2$. Between now and \euclid{}'s scheduled launch in 2019, the OGLE-IV survey observing the bulge ${\sim} 8$ months per year can detect a similar number of microlensing events to \euclid{} observing for $10$~months total. However, the \exels{} survey will be much more sensitive to low-mass planets per event.

\subsection{Photometry} \label{photometry}

In order to accurately account for the effects of severe stellar crowding on photometry of Galactic bulge stars, \mabuls{} produces simulated images for each microlensing event it simulates. The source and lens star of each microlensing event are injected into the image, with the source star's brightness updated at each epoch. Finally, relative aperture photometry is performed to measure the source brightness in each image.

The image is constructed from a smooth background component and stars drawn from the \besancon{} model catalogues. Stars are added  to the images at random locations on a fine ($9 \times 9$) sub-pixel grid, using numerical PSFs that have been integrated over pixels. Each star is tracked so that it is included at the correct position and brightness in images taken with different filters or instruments. In this way, blending is computed consistently throughout the simulation. In order to avoid small-number statistics for bright stars without using huge catalogues, we use tiered catalogues with different magnitude ranges, as listed in Table~\ref{catsize}.

At each epoch a new realization of the counts is made. Counts from stars, smooth backgrounds and instrumental backgrounds (thermal background and dark current) are Poisson realized, and fluctuations from readout noise are Gaussian realized. Photometry is performed on both the realization and a ``true'' image in a small, square, $3\times 3$ pixel aperture\footnote{Testing showed that the $3\times 3$ aperture was the optimum aperture size for simple aperture photometry in our crowded fields.} around the microlensing source with the true (input) value of the smooth background subtracted, i.e., the number of counts from stellar sources in the aperture is measured to be
\begin{equation}
N = \sum_i^{N_{\mathrm{pix}}} \left(N_{\mathrm{tot,}i} - \langle N_{\mathrm{bg}} \rangle\right),
\label{photcounts}
\end{equation}
where $N_{\mathrm{tot,}i}$ is the total number of counts in pixel $i$ and $\langle N_{\mathrm{bg}}\rangle$ is the expectation of the counts in each pixel due to all the sources of smooth background, astrophysical and instrumental. An additional gaussian fluctuation of variance $(\sigma_{\mathrm{sys}}N)^2$ is added to the final realization of the photometric measurement to simulate the effect of a systematic error floor. The photometric error is calculated as
\begin{equation}
\sigma_N^2 = \sum_i^{N_{\mathrm{pix}}} \left(N_{\mathrm{tot,}i} + \sigma_{\mathrm{read}}^2\right) + N^2\sigma_{\mathrm{sys}}^2,
\label{photerror}
\end{equation}
where $N_{\mathrm{pix}}=9$ is the number of pixels in the aperture. The $\chi^2$ of the realized photometry relative to the ``true'' photometry is the $\chi^2$ of the true model which is used to calculate the $\Delta\chi^2$ detection statistic (see the next section). Should the number of counts in a pixel (including an assumed bias level) exceed the full well depth of the detector, then the pixel saturates. If that pixel lies in the photometry aperture the data point is removed from further calculations.

It can be argued that the aperture photometry we simulate here is not appropriate for crowded fields, and that some form of PSF fitting photometry would be more realistic. While it is the case that the photometry that is performed on \euclid{} data will utilize the well known properties of the PSF to increase the photometric accuracy, it should be noted that over the small number of pixels where we perform photometry, a boxcar is a reasonable approximation of the undersampled PSF. To check that the photometric method we use does not significantly impact the number of planet detections we ran a test simulation performing photometry over a larger aperture,\footnote{The radius  of the aperture was 0.92``, covering 29 pixels. This was chosen by experimentation to optimize the photometry.} weighting the number of counts in each pixel by the intensity of the PSF in that pixel -- this weighting approximates the performance of PSF fitting photometry. For a $H$-band survey and Earth-mass planets we find that weighted photometry increases the number of planet detections by $6 \pm 2$~percent. The improvement will be larger for less massive planets and smaller for more massive planets, but in all cases will be too small to significantly affect our results. The improvement will be smaller for all other bands, because the smaller PSF in each case, and the smaller pixels on the \vis{} instrument, reduces the effect of blending. Indeed, the small under prediction of planet yields will likely be compensated by effects that we do not model in our simulations (e.g. cosmic rays or common problems that affect infrared arrays such as ghosting, charge diffusion or non-uniform pixel response functions), which are far more likely to degrade photometry than improve it.

\begin{figure*}
\includegraphics[width=\textwidth]{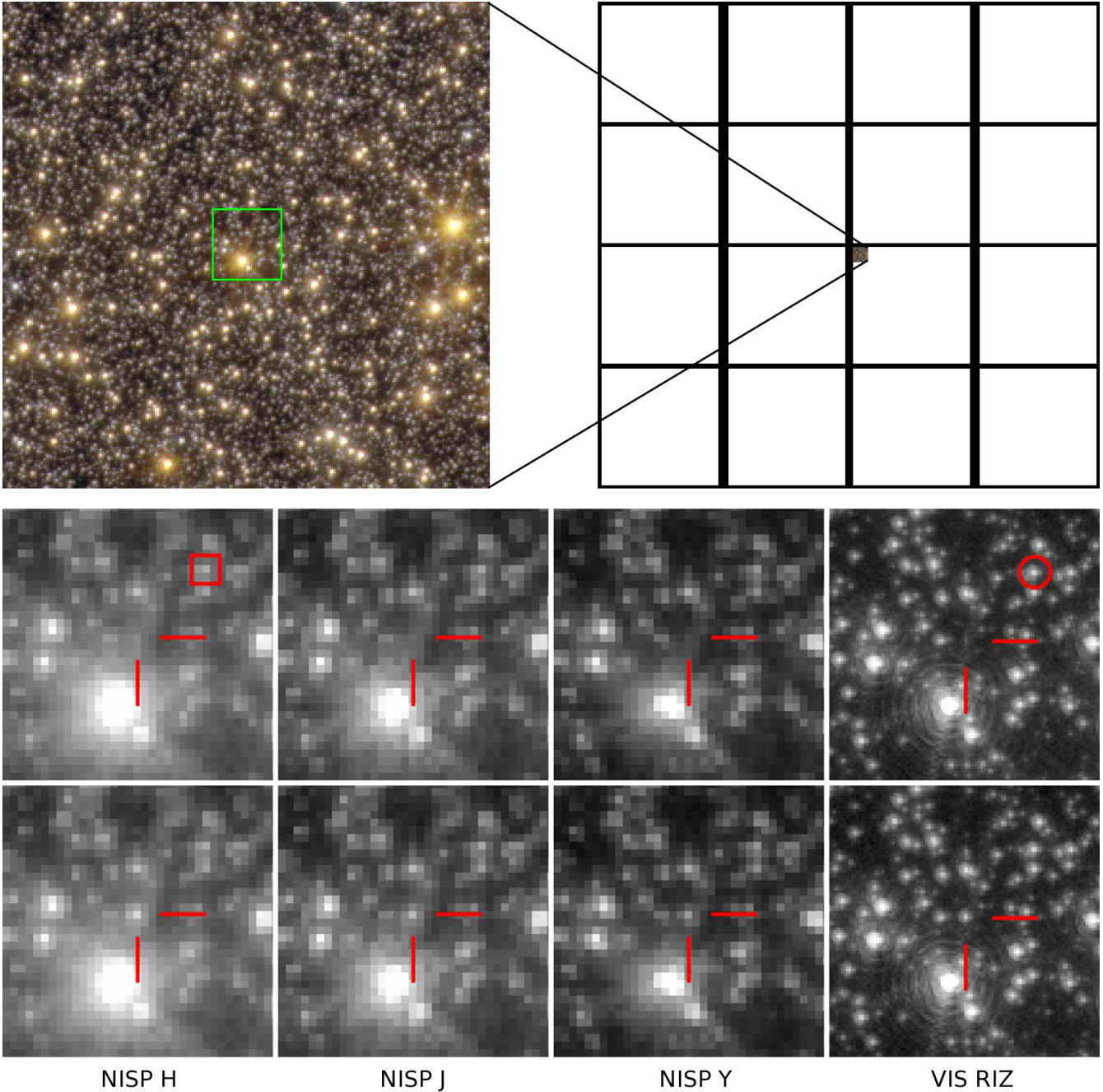}
\centering
\caption[Example of a simulated \euclid{} image]{\emph{Top left:} Example of a
  simulated false-colour composite image of a typical star-field from
  the \euclid{} \mabuls{} simulation, with colours assigned as
  red--\nisp{} $H$, green--\nisp{} $J$ and blue--\vis{} $RIZ$, each
  with a logarithmic stretch. The light green box surrounds the region
  that is shown zoomed-in in lower panels. The image covers
  $77\times77$~arcsec, equivalent to $1/64$ of a single \nisp{}
  detector, of which there are 16. These are shown to the
  right. \emph{Top right:} Approximate representation of the \nisp{}
  instrument `paw-print'. The white areas show active detector
  regions, while black areas show the gaps between detectors. In the
  corner of one of the detectors is shown the size of a simulated
  image relative to the detectors. \emph{Bottom panels:} The
  bottom panels show a small image region surrounding a microlensing
  event (located at the center and marked by cross-hairs), the top row showing images at baseline and the bottom row showing images at peak magnification
  $\mu = 28$. Panels from left to right show \nisp{} $H$, $J$, $Y$,
  and \vis{} $RIZ$ images, respectively. The small red box and red circle show the size of the aperture that was used to compute photometry in the \nisp{} and \vis{} images respectively.}
\label{imageExamples}
\end{figure*}

\begin{table}
\caption{Parameters of the
  \euclid{} telescope and detectors. Unless footnoted, all
  parameter values have been drawn from the
  \protect{\citet{redbook}}. Values in brackets are values adopted for a longer slew time of 285~sec rather than our baseline
  assumption of 85~sec. Where necessary
  parameters are explained further in the text.}
\centering
\begin{tabular}{lcccc}
\multicolumn{5}{c}{{\it Telescope parameters}}         
\\
\hline
Diameter (m)           & \multicolumn{3}{c}{} & $1.2$       
\\
Central blockage (m)   & \multicolumn{3}{c}{} & $0.4$       
\\
Slew + settle time (s) & \multicolumn{3}{c}{} & $85(285)$        
\\
\hline
\\
\multicolumn{5}{c}{{\it Detector parameters}}          
\\
Instrument & \vis{}   & \multicolumn{3}{c}{\nisp{}}         
\\
Filter     & $RIZ$  &    $Y$    &    $J$    &    $H$    
\\
\hline
Size (pixels) & $24\text{k} \times 24\text{k}$ &
\multicolumn{3}{c}{$8\text{k} \times 8\text{k}$} 
\\
Pixel scale (arcsec) & $0.1$ & \multicolumn{3}{c}{$0.3$} 
\\
PSF FWHM (arcsec) & $0.18$ & $0.3^{\ast}$ & $0.36^{\ast}$ & $0.45^{\ast}$ 
\\
&&&&\\
Bias level (e$^{-}$)   & $380^{\dag}$ & \multicolumn{3}{c}{$380^{\dag}$}
\\
Full well depth (e$^{-}$) & $2^{16}$ & \multicolumn{3}{c}{$2^{16}$} 
\\
Zero-point (ABmag) & $25.58^{\star}$ & $24.25^{\star\star}$ & $24.29^{\star\star}$ & $24.92^{\star\star}$ \\
Readout noise (e$^{-}$)& $4.5$ & $7.5^{\ast}$ & $7.5^{\ast}$ & $9.1^{\ast}$
\\
Thermal background & $0$ & $0.26$ & $0.02$ & $0.02$
\\
(e$^{-}$~s$^{-1}$) & & & & \\
Dark current (e$^{-}$~s$^{-1}$) & $0.00056^{\diamond}$ & \multicolumn{3}{c}{$0.1^{\ast}$}
\\
Systematic error & $0.001^{\dag}$ & \multicolumn{3}{c}{$0.001^{\dag}$} 
\\
Diffuse background & $21.5^{\ddag}$ & $21.3^{\ddag}$ & $21.3^{\ddag}$ &
$21.4^{\ddag}$ \\
(ABmag~arcsec$^{-2}$) & & & & \\
&&&&\\
Exposure time (s) & $540(270)$ & $90$ & $90$ & $54$ \\
Images per stack & $1$ & $3(1)$ & $3(1)$ & $5(2)$ \\
Readout time (s) & $<85$ & \multicolumn{3}{c}{$5^{\dag}$} \\
\hline
\end{tabular}\\
\begin{flushleft}
$^{\ast}$\footnotesize{\citet{Schweitzer:2010NIP}. The readout
    noise depends on the number of non-destructive reads; see text for
    further details.}\\
$^{\dag}$\footnotesize{Assumed in this work.}\\
$^{\star}$\footnotesize{M.~Cropper, private communication}.\\
$^{\star\star}$\footnotesize{G.~Seidel, private
communication}.\\
$^{\diamond}$\footnotesize{CCD203-82 data sheet, issue 2, 2007. e2v
    technologies, Elmsford, NY, USA.}\\
$^{\ddag}$\footnotesize{Calculated based on
    field locations, taking values for the zodiacal background
    from~\citet{Leinert:1998dnb}, and assuming an extra $0.2$
    magnitudes from other sources such as scattered light.}\\
\end{flushleft}
\label{parameters}
\end{table}

The properties of the detector/filter combinations that we model are listed in Table~\ref{parameters}. We note the following about the parameters listed in the
table: 
\begin{itemize}
\item The zero-point is the AB magnitude of a point source, which would
  cause one count~s$^{-1}$ in the detector, after all telescope and
  instrument inefficiencies have been accounted for. The \euclid{}
  zero-points assume end-of-life instrument
  performance~(M. Cropper, G. Seidel, private communication).
\item We distinguish between dark current and thermal background. The
  dark current is the rate of counts induced by thermal sources
  {\it within the detector pixels}, and is independent of the
  observing band. The thermal background is the count rate
  due to thermal photons emitted by all components of the spacecraft
  that hit the detector, and is therefore affected by the choice of filter.
\item For the \euclid{} simulations, we assume that the smooth
  background is due primarily to zodiacal light. To account for any additional smooth backgrounds we add an additional component with $20$~percent of the intensity of the zodiacal light. The zodiacal light background is calculated for each
  band at an elongation of 90~\degr{} in the ecliptic using data given by \citet{Leinert:1998dnb}.
\item The \vis{} $RIZ$- and \nisp{} $Y$-bands are not included in the
\besancon{} model, so we assume that the AB magnitude of a star in the
$RIZ$-band is the average of its $R$ and $I$ AB magnitudes, and
similarly we assume that the $Y$-band magnitude is the average of $I$
and $J$.
\end{itemize}

Should a pixel within the photometry aperture saturate, the data point is
flagged and is not included in the subsequent analysis. We do not 
include the effects of cosmic rays in the images, except implicitly through the use of end-of-life instrument sensitivity values. For the \euclid{} simulations, cosmic rays will
only significantly affect observations with the \vis{} instrument,
because the \nisp{} instrument, made up of infrared arrays, will use
up-the-ramp fitting with non-destructive reads~\citep{Fixsen:2000utr} to reduce readout noise and correct detector
nonlinearities~\citep{Schweitzer:2010NIP, Beletic:2008nir}. As a
consequence of the multiple reads, up-the-ramp fitting mitigates
against data loss due to cosmic rays and saturation. In order
to ensure conservatism, we assume data with saturated pixels is lost
completely. Currently we simulate the \nisp{} instrument as a
conventional CCD, but with variable read-noise determined by a
fundamental read-noise ($13$~e$^{-}$) and the number of
non-destructive reads during an exposure, which we assume occur at a
constant rate of once every ${\sim} 5$~s~\citep{Schweitzer:2010NIP}. We do not currently 
simulate the more complicated effects of  charge smearing~\citep[see, e.g.,][]{Cropper:2010VIS} and ghosts from
bright stars.

\begin{figure}
\includegraphics[width=\figuresize]{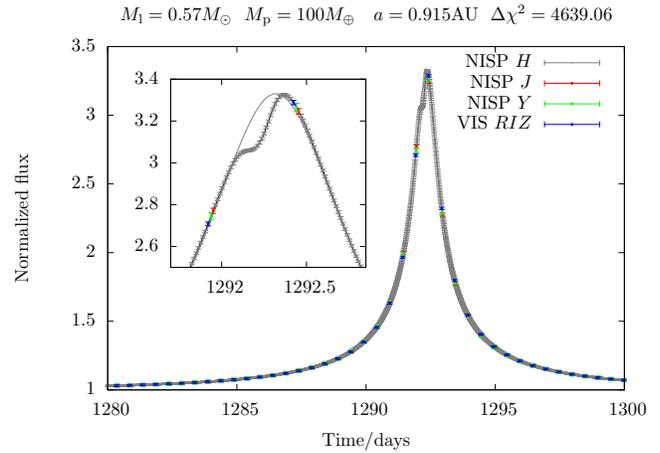}
\caption[Lightcurve of the event Figure~\ref{imageExamples}]{Lightcurve of the simulated event shown in
  Figure~\protect{\ref{imageExamples}}. Fluxes are plotted normalized to the baseline and blending in the $H$-band. Grey, red, green and blue show data from \nisp{} $H$, $J$, $Y$ and
  \vis{} $RIZ$, respectively. The
  event reaches a peak magnification of ${\sim} 28$, but the
  normalized flux only increases by a factor of ${\sim} 3.3$ because the
  source ($H_{\mathrm{AB}}=20.9$) is blended with the diffraction spike of a much brighter star about $10$ \nisp{}-pixels away, and several other fainter stars, including the lens ($H_{\mathrm{AB}}=32.0$). At baseline the source contributes just $8$~percent of the total flux in the $H$-band photometry aperture, though in the $RIZ$-band aperture it contributes about $18$~percent. Some of the event parameters are shown above the figure: $M_{\mathrm{l}}$ is the host-star mass; $\Delta\chi^2$ is introduced in the next section. The inset shows the peak of the event, where a planetary signature is clearly detected, relative to the single-lens lightcurve (the grey line) that would have been observed were the planet not present. Data points are not scattered for clarity.}
\label{imageLC}
\end{figure}

For the \euclid{} simulations we use numerical PSFs computed for each
instrument and each band. The \nisp{} PSFs are computed near the edge
of the detector field of view and include the effect of jitter and
instrument optics in the worst case scenario~(G.~Seidel, private
communication). The \vis{} PSF is similarly computed~(M. Cropper,
private communication). Figure~\protect{\ref{imageExamples}} shows an 
example of a simulated, colour-composite image of a field with a
microlensing event at its centre. The very brightest, reddest stars in the image are bright bulge giants of ${\sim} 1$~solar mass and ${\sim} 80$--$120$ solar radii. The much more numerous, but still bright and red, stars are red-clump giants in the bulge; bluer stars of a similar brightness are main-sequence F-stars in the disc. The fainter, resolved
stars are turn-off and upper-main-sequence stars in the
bulge. The figure also shows an approximate representation of the
scale of the \nisp{} instrument, which is constructed from $4 \times
4$ HgCdTe infrared arrays, each of $2048 \times 2048$~pixels covering
$10 \times 10$~arcmin, for a total detector area of $0.47$~deg$^2$; the gaps between detectors are
approximately to scale. We do not include these gaps in the simulation
and assume the instrument is a single $8192\times 8192$-pixel
detector. The lower section of
Figure~\ref{imageExamples} shows a set of
zoomed-in image sections, centred on the microlensing event at peak
and at baseline, in each of the \nisp{} and \vis{}
bands. Note the diffraction spikes and Airy rings in the \vis{}
images; such spikes and rings can
significantly affect photometry of faint sources. Figure~\ref{imageLC}
shows the lightcurve of the simulated event with peak magnification $\mu = 28$ that occurs in the example
image, including the points that are lost to saturation. For the sake of
computational efficiency only a small image segment, just bigger than
the largest aperture, is simulated in the standard operation of \mabuls{}.

\subsection{Planet detections}

To determine whether a bound planet is detected in a microlensing event we
use three criteria, which will be further explained below:
\begin{enumerate}
\item the $\Delta\chi^2$ between the best-fitting single-lens model and the best-fitting planetary model must be greater than $160$,
\item the $\Delta\chi^2$ contribution from the primary observing band must be at least half of the total $\Delta\chi^2$,
\item the time of closest approach between the lens star and the source ($\tzero$) must be within one of the $30$-day observing seasons.
\end{enumerate}

For the first criterion, we assume the best-fitting planetary model to be the true underlying model that was used to simulate the event. We choose $\Delta\chi^2>160$, which corresponds to a $\sigma>12.6$ detection of the planet, because we find that the signatures of low-mass planets at this level of significance can usually be seen as systematic deviations from a single-lens lightcurve by eye (see e.g. event (c) in Figure~\ref{lcexamples} below). This is in contrast to \citet{Gould:2010pps}, who choose a higher threshold $\Delta\chi^2>500$ for planets in high-magnification microlensing events. \citet{Gould:2010pps} were analyzing data from multiple, small ground based observatories, which can suffer from various systematic effects (e.g. due to weather, differences in instrumentation, atmospheric effects in unfiltered data, etc.) that make the accurate estimation of photometric uncertainties, and hence also $\chi^2$, extremely difficult. More recent work by \citet{Yee:2012b} suggests that while a threshold of $\Delta\chi^2>500$ may be appropriate for planets in high-magnification events, a lower threshold of $\Delta\chi^2 \gtrsim 200$ is likely to be more appropriate for ground-based detection of planets in standard microlensing events, where the planetary signal is less ambiguous than in high-magnification events. A space-based microlensing data set will be much more uniform than ground-based data and will have much better characterized systematic effects, especially in the case of \euclid{}, whose design is driven by difficult, systematics-limited, weak lensing galaxy shape measurements. In order to see if planetary parameters could be measured from $\Delta\chi^2\approx 160$ lightcurves, we fitted a few simulated lightcurves using a Markov Chain Monte Carlo minimizer and found that even with $\Delta\chi^2\approx 100$ it was still possible to robustly measure the basic microlensing event parameters, including the mass ratio, separation and in events where it was important, the source radius crossing time \citep[see Appendix A of ][]{Penny:2011thesis}. 

\begin{figure}
\includegraphics[width=\figuresize]{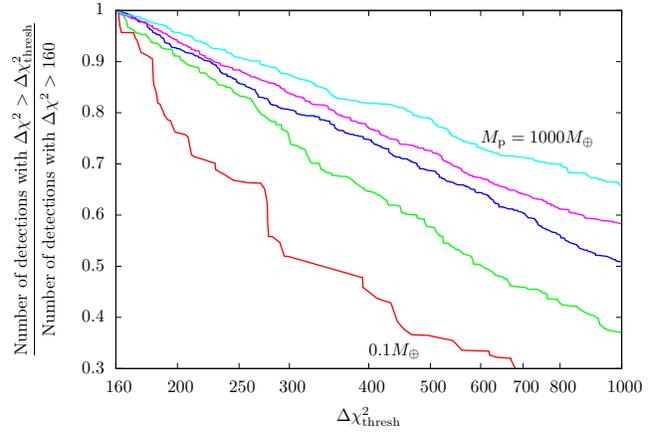}
\caption{The effect of changing the $\Delta\chi^2$ threshold on the number of planet detections. The number of planet detections with a $\Delta\chi^2$ threshold $\Delta\chi_{\mathrm{thresh}}^2$, relative to the number of detections with $\Delta\chi^2>160$, is plotted against $\Delta\chi_{\mathrm{thresh}}^2$. Red, green, blue, magenta and cyan lines show the number of detections for $0.1$-, $1$-, $10$-, $100$- and $1000$-$\mearth$ planets, respectively.}
\label{chi2thresh}
\end{figure}

Our choice of $\Delta\chi^2>160$ also aids comparison with other simulations which have chosen the same threshold~\citep[Gaudi et al., unpublished]{Bennett:2002ssm}, and is also the value adopted by the \wfirst{} science definition team for their calculations of the exoplanet figure of merit~\citep{wfirstir}. Despite the widespread adoption of $\Delta\chi^2>160$ as a threshold for planet detections in space-based microlensing surveys, it is worth considering the effect of changing the threshold. Figure~\ref{chi2thresh} plots the number of detections with $\Delta\chi^2$ greater than a variable threshold $\Delta\chi_{\mathrm{thresh}}^2$, relative to our chosen threshold of $160$. A higher $\Delta\chi^2$ threshold of $\Delta\chi_{\mathrm{thresh}}^2=200$ would reduce yields by only ${\sim}25$~percent for Mars-mass planets and less than ${\sim}10$~percent for higher mass planets. Even an extremely conservative threshold $\Delta\chi_{\mathrm{thresh}}^2=500$, such as used by \citet{Gould:2010pps} for ground-based observations, reduces detections by $40$--$20$~percent, depending on planet mass, above $1\mearth$. Such a reduction in yield would not prevent \euclid{} from probing the abundance of Earth-mass planets, but may significantly affect measurements for Mars-mass planets. However, such an extremely conservative cut will almost certainly not be necessary. 

Returning to the definition of selection criteria, the second criterion is chosen in order to allow fair comparisons between the different bands that \euclid{} can observe in. By requiring that the contribution to $\Delta\chi^2$ from the primary observing band is at least half of the
total $\Delta\chi^2$, we ensure that the primary band
provides most of the information about the planet, and do not count as detections events where a planet is detected but most of the data are lost (e.g., due to
saturation) or provides little information.

The final criterion is chosen to increase the chance that the microlensing event timescale is well constrained. The season length for microlensing observations on \euclid{} will be short, ${\sim}30$ days, due to the restrictions of the spacecraft's sunshield. This can result in only a small portion of longer timescale events being monitored, and may also mean that the event timescale can not be constrained. To first order, it is the ratio of the timescale of the planetary perturbation to the timescale of the main microlensing event which is used to measure the planetary mass ratio. Without the denominator of the ratio, the planetary mass ratio, and hence planetary mass cannot be estimated. Note however that it may be possible to constrain the event timescale from the curvature of the lightcurve without the peak, as in event (d) shown in Figure~\ref{lcexamples} below.

\begin{figure*}
\subfigure[]{\includegraphics[width=\figuresize]{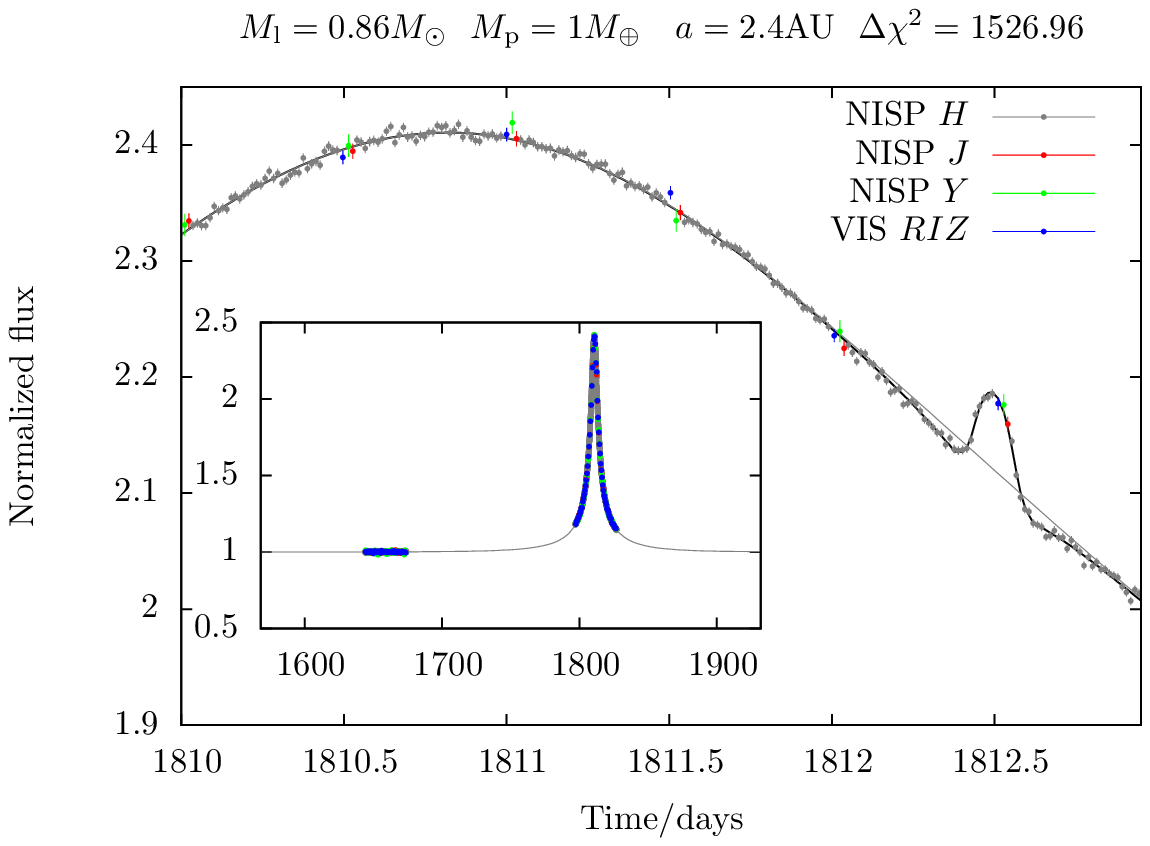}}
\subfigure[]{\includegraphics[width=\figuresize]{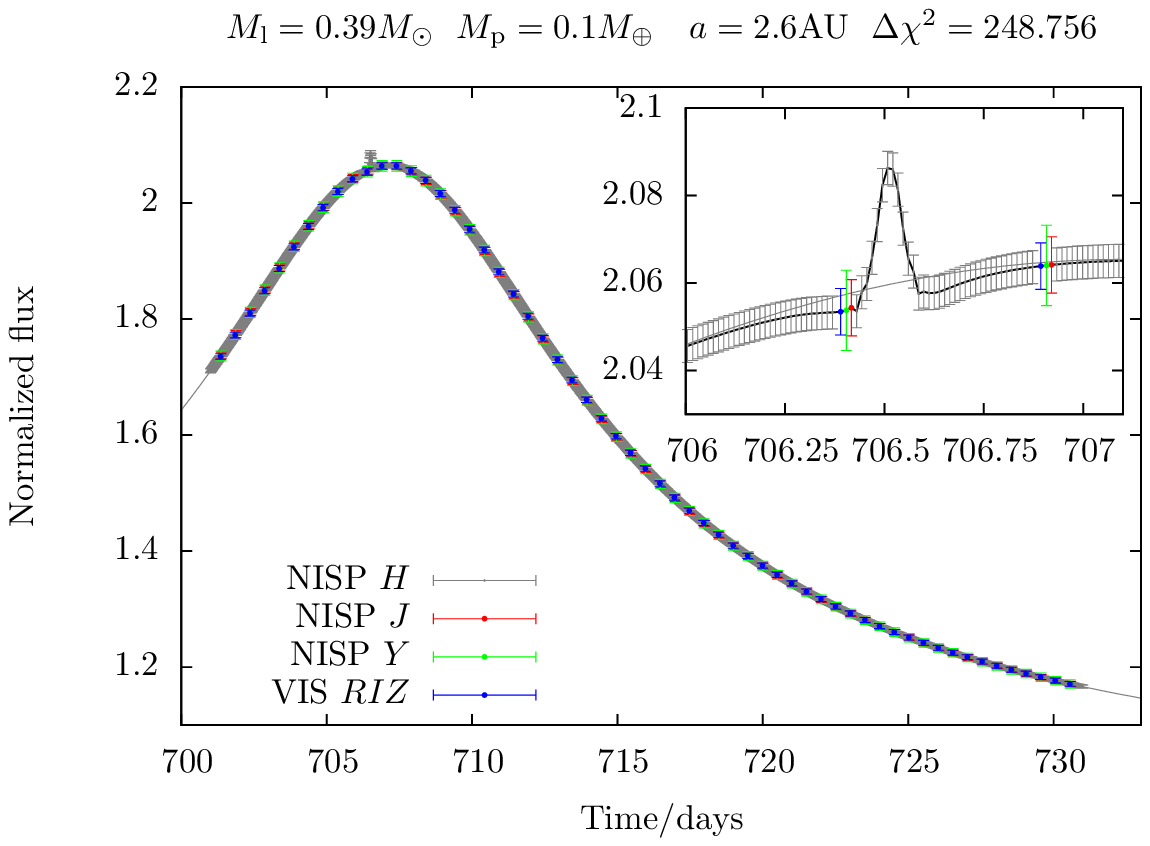}}\\
\subfigure[]{\includegraphics[width=\figuresize]{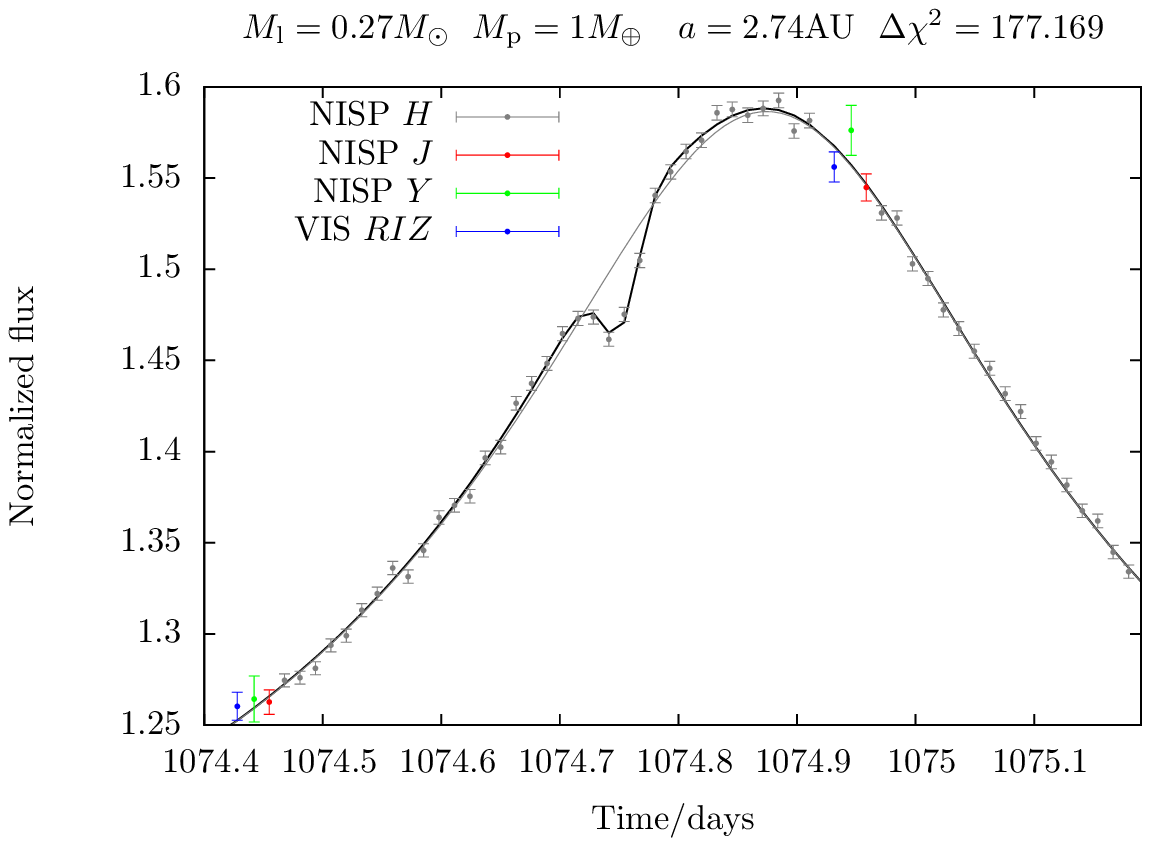}}
\subfigure[]{\includegraphics[width=\figuresize]{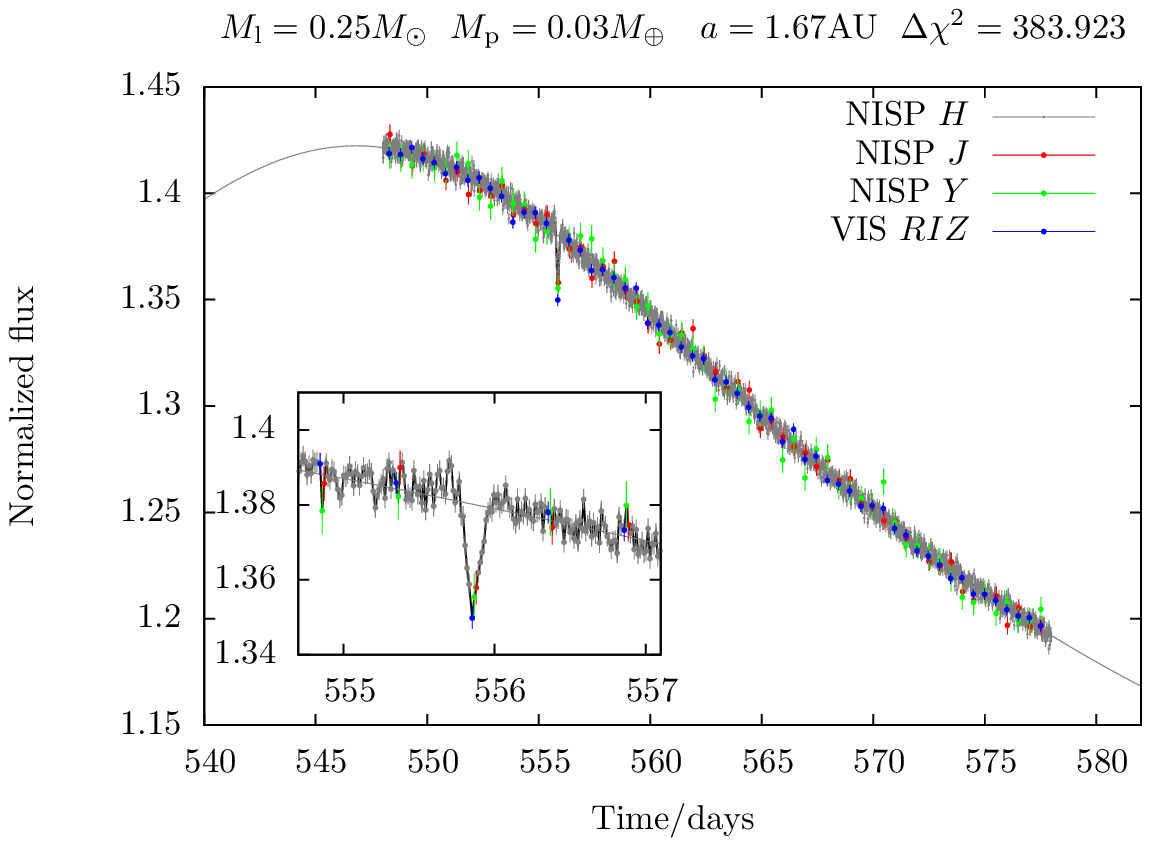}}\\
\subfigure[]{\includegraphics[width=\figuresize]{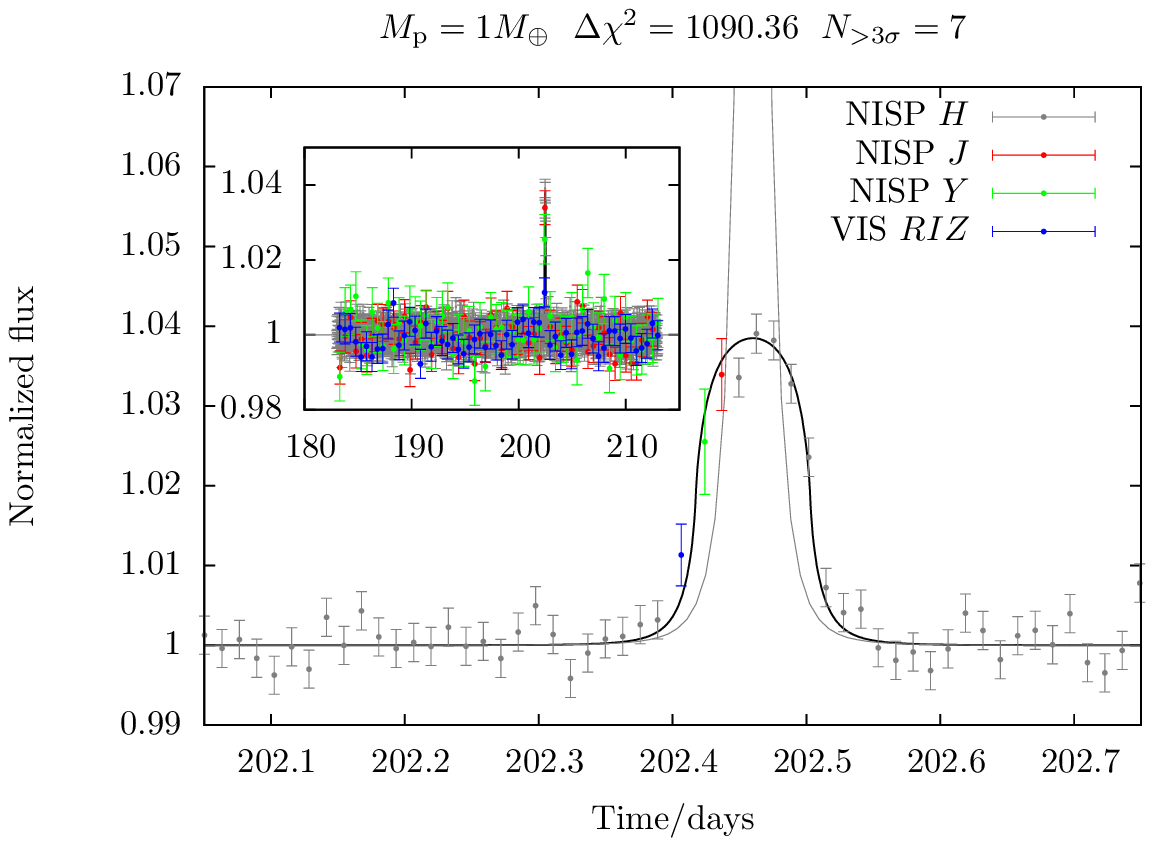}}
\subfigure[]{\includegraphics[width=\figuresize]{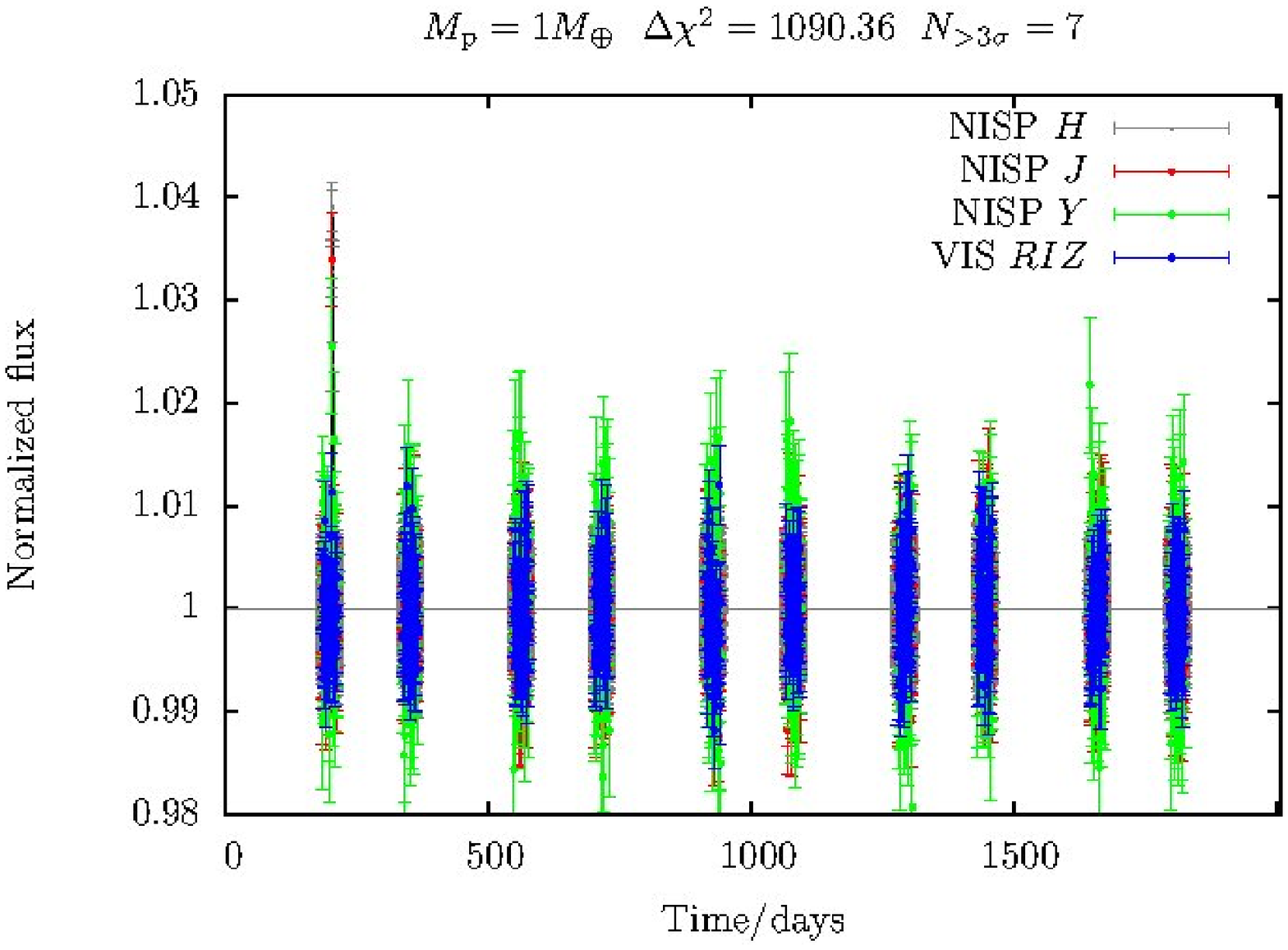}}
\caption[Example lightcurves]{Example lightcurves from the \mabuls{}
  simulation of \exels{}. The left column shows Earth-mass planet detections, with (a) showing a strong detection, (c) showing a detection close to the $\Delta\chi^2$ threshold, and (e) showing an Earth-mass free-floating planet detection. (f) shows the full lightcurve of the free-floating planet detection in (e). The lightcurve in (b) shows a Mars-mass planet detection, but with the data points not scattered about the planetary lightcurve in order to emphasize the relative sizes of the photometric error bars. The lightcurve in (d) shows a $0.03$-$\mearth$ planet that causes a signal well above our $\Delta\chi^2$ threshold, but which is not counted as a detection because we require that the time of lens-source closest approach (the peak of the primary lensing event) be within an observing season. The inset figures, where included, either zoom in on planetary features or zoom out to show a larger section of the lightcurve. Grey, red, green and blue points with error bars show the simulated photometric data, while the black line shows the true lightcurve and the grey line shows the point-source single-lens lightcurve that would be seen if the planet were not present. In (e) the grey curve shows the lightcurve that would be seen if the source were a point, not a finite disc as is actually the case. In each lightcurve the flux has been normalized to the $H$-band flux, taking into account blending. The host mass, planet mass and semimajor axis, and $\Delta\chi^2$ are shown above each lightcurve.}
\label{lcexamples}
\end{figure*}

Figure~\ref{lcexamples} shows some example lightcurves from the
simulation. The lightcurves show planet detections with varying
degrees of significance, ranging from a detection that narrowly passed
the $\Delta\chi^2$ cut (lightcurve (c), $\Delta\chi^2=177$) to a very significant
detection (lightcurve (a), $\Delta\chi^2=1527$). Note however, that
many events will have much higher $\Delta\chi^2$ than this, up to
$\Delta\chi^2\approx 10^{6\text{--}7}$. The example lightcurves also cover a
range of host and planet masses; the event with the lowest-mass planet
is event (d), which has a planet mass $\mpl=0.03\mearth$ and detected
with $\Delta\chi^2=384$; however, due to our second criterion that $\tzero$ must lie in an observing season, this event is not counted as a detection. 

\subsubsection{Free-floating planets}

To determine the expected number of free-floating planet detections we
adopt similar detection criteria to those of \citet{Sumi:2011ffp}. We
require that in order to be classed as a detection, a free-floating
planet lightcurve must have:
\begin{enumerate}
\item at least $6$ consecutive data points (in any band) detected at greater than $3\sigma$
  above baseline; and
\item $\Delta\chi^2 > 500$ relative to a constant baseline model, using all the
  data points in the primary observing band that satisfy the first
  criteria.
\end{enumerate}
These criteria are in fact far more stringent than the corresponding
criteria imposed by \citet{Sumi:2011ffp}, but we chose them to remain
conservative, as we do not impose other criteria relating to the
quality of microlensing model fits and images that
\citet{Sumi:2011ffp} use.

\section{Expected yields} \label{yields}

In this section we discuss the results of the \mabuls{} simulations of \exels{}. Unless otherwise noted, we present the results assuming that each lens star in the simulation is orbited by a single planet of mass $\mpl$ with semimajor axis in the range $0.03<a<30$~AU.

\begin{figure}
\includegraphics[width=\figuresize]{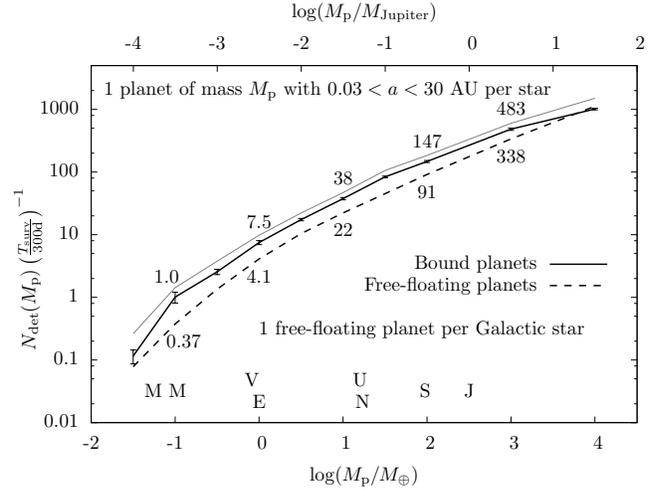}
\caption[Expected planet yield for \euclid{}]{Number of planets
  detected in a $300$-day survey by \euclid{}, plotted against planet mass $\mpl$. The survey is primarily conducted in the \nisp{}
  $H$-band. The solid black line shows the expected bound planet yield,
  assuming one planet of mass $M_{\mathrm{p}}$ per star with semimajor
  axis $0.03\le a< 30$~AU; error bars show our estimated statistical errors from simulations of a finite number of lightcurves. The solid grey line shows the yield if the third cut on the time of the event peak is not applied. The dashed line shows the expected yield of free-floating planets, assuming there is $1$ free-floating planet per Galactic star. The masses of Solar System planets are
  indicated by letters, and the numbers above/below the lines show the yields when applying the full sets detection criteria.}
\label{Ndetections}
\end{figure}

Figure~\ref{Ndetections} shows the expected number of planet
detections plotted against planet mass, using a naive
assumption that there is one planet of mass $\mpl$ and semimajor axis
$0.03<a<30$~AU per star. The error bars on this plot, and all subsequent
plots of the yield, show the uncertainty due to the finite number of
events that we simulate. Error bars are not shown for the
free-floating planet simulation as they are similar to or smaller than the line
thickness. For this naive assumption we expect a \euclid{}
planetary microlensing survey would detect ${\sim} 8$, $38$ and $147$ bound Earth-, Neptune- and Saturn-mass planets (within $1$-decade wide mass bins), and roughly half as many free-floating planets of the same masses. \euclid{} is sensitive to planets with masses as low as $0.03\mearth$, but the detection rate 
for such low-mass planets is likely to be small unless the exoplanet mass function rises steeply in this mass regime. 

\begin{figure}
\includegraphics[width=\figuresize]{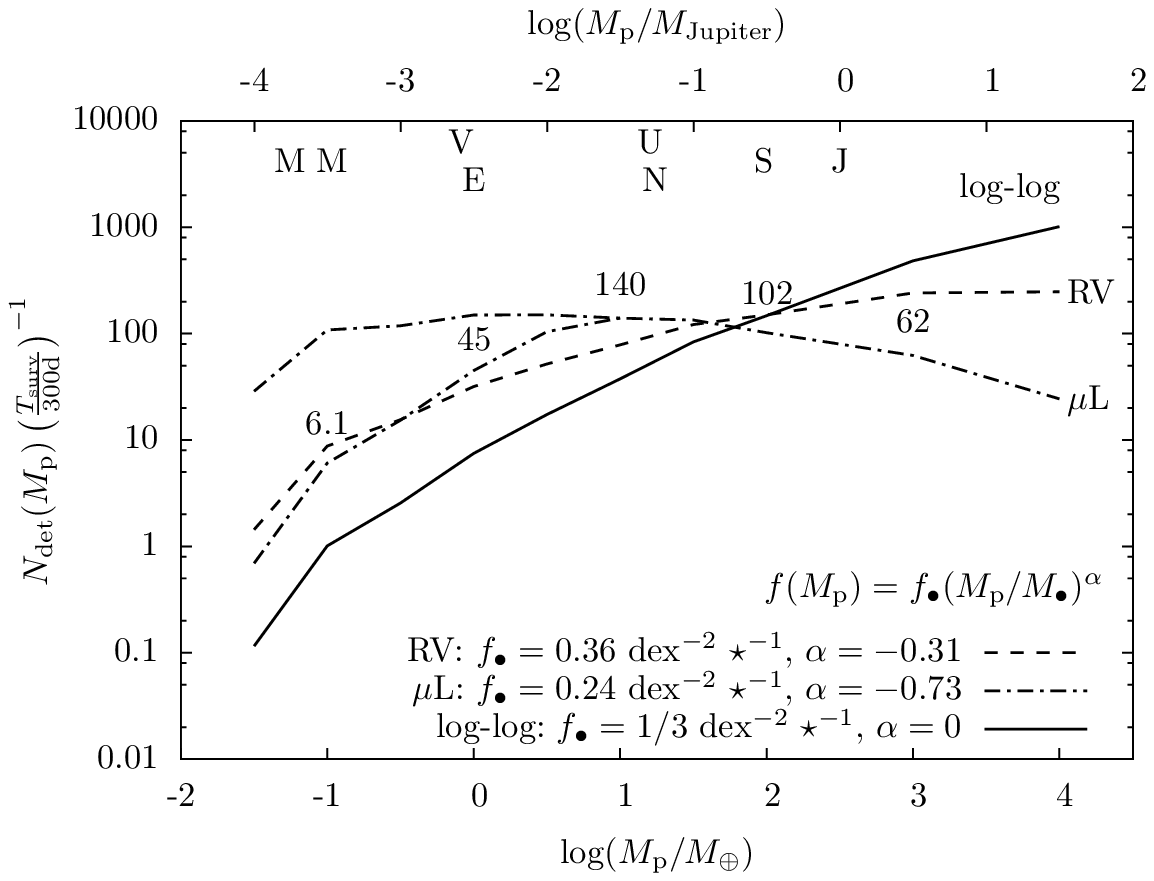}\\
\includegraphics[width=\figuresize]{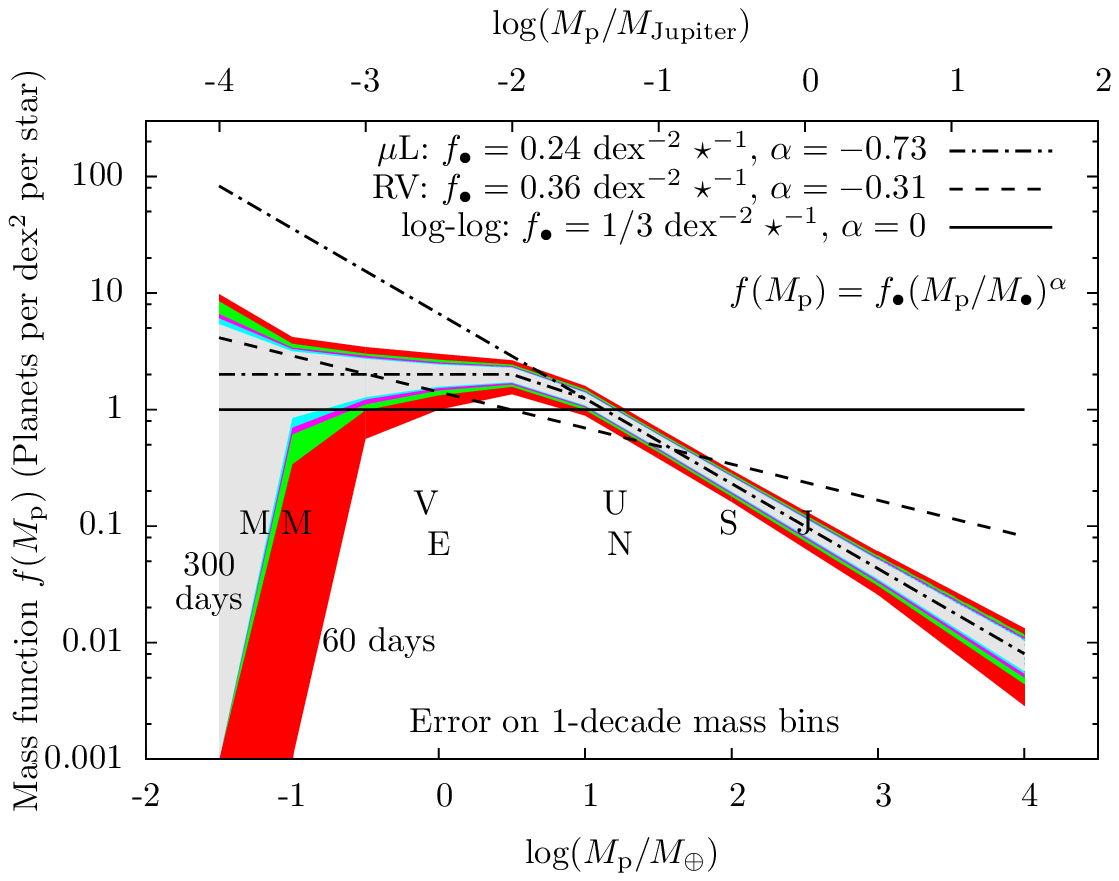}
\caption[Planet yield for empirical mass functions]{The upper panel shows predictions of the planet yield based on recent estimates of
  the planet abundance and planet-mass distribution. The solid line
  shows a naive logarithmic prior of one planet per decade of mass and
  semimajor axis per star. The dashed line (labelled RV) shows the
  expected yield using an extrapolation of the mass-function slope
  measured by \citet{Cumming:2008mpd} using RV data combined with 
  a normalization measured by \citet{Gould:2010pps} from microlensing
  data, which \citet{Gould:2010pps} argue is compatible with the slope of the
  semimajor axis distribution found by \citet{Cumming:2008mpd}.
  The dot-dashed line (labelled $\mu$L) shows the expected yield using the mass function slope and normalization measured from microlensing data by
  \citet{Cassan:2012pmf}. A branching dot-dashed line, and the numbers above it, show the yield if the \citet{Cassan:2012pmf} microlensing mass function saturates at $2$ planets per dex$^2$ per star. The lower panel shows the actual form of each of the mass functions shown in the top panel. The filled, coloured regions show the size of model-independent $1$-$\sigma$ statistical (square root $N$) errors on measurements of the planet abundance in $1$-decade mass bins centred at $\mpl$, assuming the saturated microlensing mass function and also assuming that only half of the planet detections have host mass measurements. The red, green, magenta, cyan and grey regions show the error bars for \euclid{} microlensing surveys lasting $60$, $120$, $180$, $240$ and $300$ days respectively. This implies a $300$-day \euclid{} microlensing survey would measure the abundance of Earth-mass and Mars-mass planets to approximately $4.7$ and $1.7$-$\sigma$ respectively, whereas a $120$-day \euclid{} survey would reach just $3.0$ and $1.2$-$\sigma$ significance, both assuming the saturated mass function.}
\label{Ntruedetections}
\end{figure}

Recent measurements of planet abundances using several techniques have
shown that the often used logarithmic planet mass function prior is
quite unrealistic. Multiple studies have suggested that the
number of planets increases with decreasing planet mass~\citep{Cumming:2008mpd, Johnson:2010cps, Sumi:2010nps,
  Howard:2012, Mayor:2011hor, Cassan:2012pmf} and that planets
are not distributed logarithmically in semimajor axis~\citep{Cumming:2008mpd}. This picture is also supported by planet population synthesis
models~\citep{Mordasini:2009pps,Mordasini:2009sco,Ida:2008ilp}. In Figure~\ref{Ntruedetections} we consider a more realistic two-parameter power-law
planetary mass function:
\be
f(\mpl) \equiv \frac{\dd^2 N}{\dd\log\mpl \dd\log a}= \fpivot \left(\frac{\mpl}{\mpivot}\right)^{\alpha},
\label{massfunction}
\ee
where $f(\mpl)$ is now the number of planets of mass $\mpl$ per decade
of planet mass per decade of semimajor axis per star and where $\fpivot$ is the
planet abundance (in dex$^{-2}$~star$^{-1}$) at some mass $\mpivot$ about which the mass function pivots. Here, $\alpha$ is the slope of the mass function, with
negative values implying increasing planetary abundance with
decreasing planetary mass. For simplicity, and because there are no
measurements of the slope of the planetary semimajor axis
distributions in the regime probed by microlensing, we assume that
$\dd N/\dd \log a$ is constant. 

We use two estimates of the mass-function
parameters based on measurements made using both RV and microlensing
data sets. The first, more conservative mass function (in terms of the yield
of low-mass planets) uses the mass-function slope $\alpha = -0.31 \pm
 0.20$ measured by \citet{Cumming:2008mpd} from planets with periods
in the range $T=2$--$2000$~d, detected via radial velocities. For the
normalization we use $\fpivot=0.36 \pm 0.15$ at $\mpivot \approx
80\mearth$, measured by \citet{Gould:2010pps} from high-magnification
microlensing events observed by MicroFUN. \citet{Gould:2010pps} argue
that this value is consistent with the abundance and semimajor axis
distribution measured by \citet{Cumming:2008mpd}, 
extrapolated to orbits with $a\approx 2.5$~AU. We note that the host stars studied by
\citet{Cumming:2008mpd} typically have higher masses than those that
are probed by microlensing. We call the combination of the \citet{Cumming:2008mpd} slope and \citet{Gould:2010pps} normalization,
the RV mass function.
The second mass function we consider has a mass function slope $\alpha=-0.73 \pm 0.17$ and normalization $\fpivot=0.24_{-0.10}^{+0.16}$ at $\mpivot=95\mearth$, as measured by \citet{Cassan:2012pmf} from microlensing detections. We call this the microlensing ($\mu$L) mass function. We note that at low masses, the extrapolation of the
microlensing mass function implies close packing of planetary
systems. We also plot the microlensing mass function assuming that it saturates at a planet abundance of
$2$~dex$^{-2}$~star$^{-1}$. However, we note that the Kepler~20 planetary system comprises five exoplanet candidates so far \citep{Gautier:2012}, all within about 1~dex in both mass and separation. Our saturation limit is therefore likely to be conservative.

\begin{table}
\caption{Expected total number of planet detections by a $300$-day \euclid{} microlensing survey for different mass functions (with planet masses in the range $0.03<\mpl/\mearth<3000$ (roughly $0.6$ Mercury-mass to $10$ Jupiter-mass).}
\begin{tabular}{@{\extracolsep{\fill}}lr}
\hline
Mass function & Number of detections\\
\hline
log-log & 718 \\
RV & 502 \\
$\mu$L & 541 \\
$\mu$L saturated & 356 \\
\hline
\end{tabular}
\label{totalDetections}
\end{table}

Figure~\ref{Ntruedetections} plots the yields that would be expected
for the different mass functions. Perhaps the most important thing that the top panel of Figure~\ref{Ntruedetections} highlights is that, despite the degree of uncertainty in the extrapolation to low planet masses provided by empirical estimates of the mass functions, we can expect a $300$-day \euclid{} survey to detect significant numbers of planets of Mars-mass and above. Table~\ref{totalDetections} shows the total number of detections expected for each mass function. The number of expected detections imply that \euclid{} data would allow the different model mass functions to be discriminated between. In fact, we can look at the power of \euclid{} to measure the mass function more easily in the lower panel of Figure~\ref{Ntruedetections}.

The lower panel of Figure~\ref{Ntruedetections} shows the expected uncertainty on the planet abundance in one-decade mass bins, assuming the saturated microlensing mass function and that half the \euclid{} planet detections have mass measurements. Such mass measurements can be made by estimating the mass of the host from photometry of the host star. Such an estimate should be possible for many of the hosts using ExELS survey data alone, thanks to the extremely deep, high-resolution images that can be built by combining the randomly dithered survey images. A stack of such images, one for each season, will allow the light of the source, host and any unrelated stars to be disentangled as they separate due to their mutual proper motions. \citet{Bennett:2007phc} give a detailed discussion of how these mass measurements are made. \citet{Bennett:2007phc} estimate that such mass measurements should be possible in most space-based planetary microlensing detections. However, it may be the case that the larger pixels of the \nisp{} instrument preclude full photometric host-mass measurements, but even if this is so, the deep, high-resolution images from the \vis{} channel should provide constraints. Even without mass measurements, Figure~\ref{Ntruedetections} indicates the precision of measurements of the mass-{\it ratio} function, which would encode much of the same information. The uncertainties shown by the coloured bands in the figure are the uncertainty on the absolute abundance of planets in each bin. This is in contrast to measurements such as those of \citet{Cumming:2008mpd} and \citet{Cassan:2012pmf}, which are the uncertainties on a small number of power-law model parameters {\it assuming} the models are correct. If we here assume that the saturated microlensing mass function is correct, then we can see that a $300$-day \euclid{} microlensing survey would measure the abundance of Earth mass planets to be $2$ per star with a significance of $4.7$-$\sigma$, and similarly the abundance of Mars mass planets to $1.7$-$\sigma$. However, if the microlensing program were only $120$ days, the significance of the abundance measurements would reduce to $3.0$- and $1.2$-$\sigma$ for Earth- and Mars-mass planets, respectively.

\subsection{The $\mpl$--$a$ diagram}

\begin{figure}
\includegraphics[width=\figuresize]{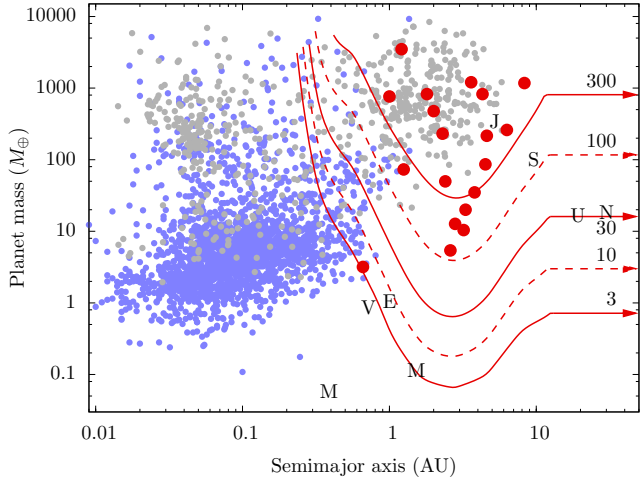}
\caption{The sensitivity of \euclid{} in the $\mpl$--$a$ plane. Red lines
  show the expected yield of a $300$-day \euclid{} survey with $60$
  days of observations per year, plotted against
  planet mass and semimajor axis, assuming one planet per star at each
  point in the planet mass--semimajor axis plane.
Horizontal
  arrows are plotted when the expected yield of free floating planets
  of that mass exceeds the yield of bound planets (assuming one free
  floating planet per star). 
  The grey points show planets listed by the Exoplanets Orbits Database as of 17th
  March 2012~\citep{exoplanetsorg}, and light blue points show candidate planets
  from the \kepler{} mission~\citep{Batalha:2013}, with masses calculated using the
  mass-radius relation of \citet{Lissauer:2011kmp}. The red points show
  planets detected via microlensing to date.}
\label{sensitivity}
\end{figure}

We have discussed the ability of our simulated survey to probe the
planetary mass function, but a perhaps more important goal of such a
survey is to explore the planet mass--semimajor axis ($\mpl$-$a$)
plane where planet formation models predict a lot of
structure~\citep[e.g.,][]{Ida:2004dpf, Mordasini:2009pps}. Figure~\ref{sensitivity}
plots contours of planet detection yields for the simulated survey in
the $\mpl$-$a$ plane, assuming there is one planet per host at a
given point in the plane. 
The positions of planet detections to date, by all detection methods (RV,
transits, direct detection, timing and microlensing) are also shown,
as well as candidate planets detected by
\kepler{}~\citep{Batalha:2013}, which have been plotted by assuming
the planetary mass-radius relation, $\mpl = (\rpl/\rearth)^{2.06}
\mearth$, which is used by \citet{Lissauer:2011kmp}. It is clear
that microlensing surveys probe a different region of the $\mpl$-$a$
plane to all other detection methods, covering planets in orbits ${\sim}
0.2$--$20$~AU, as well as free floating planets. Note that
microlensing can be used to detect planets with any semimajor axis
larger than ${\sim} 20$~AU, but there is a significant chance that the
microlensing event of the host will not be detectable. These cases
will be classified as free-floating planet detections~\citep[see
  e.g.][]{Sumi:2011ffp, Bennett:2012}. The peak sensitivity of the simulated \euclid{} survey is
at a semimajor axis $a\approx 1.5$--$5$~AU, in good agreement with previous
simulations of space-based microlensing surveys~\citep[Gaudi et al., unpublished]{Bennett:2002ssm}. The planets
to which \euclid{} is sensitive lie in wider orbits than those detectable by
\kepler{}, and stretch to much lower masses than can be detected by RV in
this semimajor axis range, reaching down to Mars mass. The range of
semimajor axis probed by \euclid{} decreases with decreasing mass,
from ${\sim}0.2$ to more than $20$~AU for Jupiter-mass planets, down to
${\sim}1$--$14$~AU for Earth-mass planets and ${\sim}1.5$--$5$~AU for Mars-mass
planets. There will be a significant degree of overlap between
\euclid{} and full-mission \kepler{} detections at separations
$0.3\lesssim a \lesssim 1$~AU. Similarly, at masses larger than
$\mpl\gtrsim 50\mearth$, there will be overlap with RV surveys over a
wide range of semimajor axes. Both overlaps will facilitate
comparisons between the data sets of each technique. It should be
noted however, that the host populations probed by each technique are
different, as we will see in the next section.

\begin{figure}
\includegraphics[width=\figuresize]{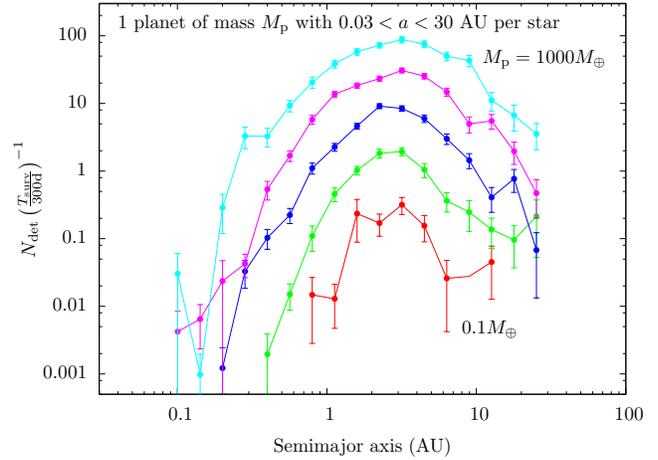}
\caption[Planet yield as a function of $a$]{Predictions of the planet yield as a function of semi-major axis $a$. 
The red, green, blue, 
magenta and cyan lines denote yields for 0.1,1,10,100 and 1000~$\mearth$, respectively.}
\label{Nva}
\end{figure}

Figure~\ref{Nva} plots the expected yield for various planet masses
as a function of semimajor axis $a$, using a simplistic assumption of one planet per host at the given mass and separation. The peak sensitivity of \euclid{} is to planets with
semimajor axis $a\approx 1.5$--$5$~AU. The sensitivity is ${\sim}10$~percent of the peak sensitivity in the range $0.5 \lesssim a \lesssim
20$~AU. 

\begin{figure}
\includegraphics[width=\figuresize]{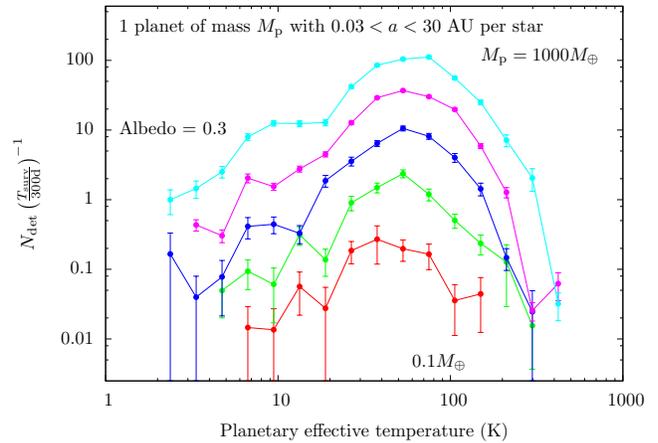}
\caption{The number of planet detections plotted against the planetary
  effective temperature, assuming an albedo of 0.3. Lines are as for Figure~\protect{\ref{Nva}}.}
\label{Teff}
\end{figure}

Figure~\ref{Teff} plots the distribution of planet detections as a
function of the effective temperature of the planet, calculated as
\be
T_{\mathrm{eff,p}} =
\sqrt{\frac{R_{\mathrm{l}}}{2a}}\left(1-A\right)^{1/4}T_{\mathrm{eff,l}},
\label{teff}
\ee
where $R_{\mathrm{l}}$ is the radius of the host star, $A$ is the
planet's albedo, assumed to be $A=0.3$ and $T_{\mathrm{eff,l}}$ is the
effective temperature of the star. Both $R_{\mathrm{l}}$ and
$T_{\mathrm{eff,l}}$ are provided as outputs of the \besancon{}. The distribution of detected planet temperatures peaks at $\sim
50$--$80$~K, with a long tail towards lower temperatures and a rapid decline
towards higher temperatures. However, there should still be a
small number of detections of planets with effective temperatures
$\gtrsim 200$~K. 

\subsection{Host star properties}

\begin{figure}
\includegraphics[width=\figuresize]{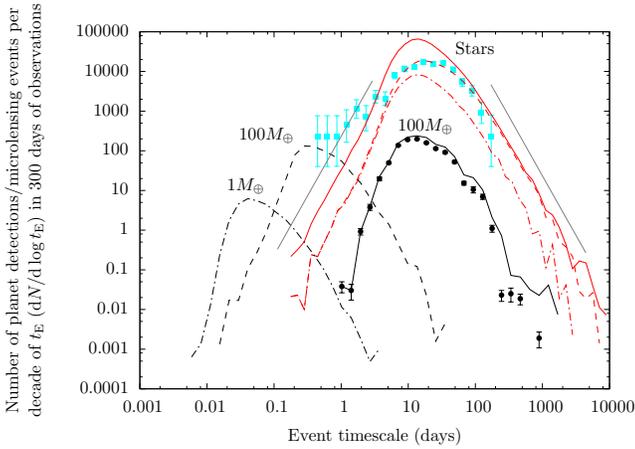}
\caption{The distribution of microlensing timescales. Red curves show the distribution of event timescales for stellar microlensing events. The solid curve shows all events with impact parameter $\uzero\le 1$, regardless of whether they are detected, the dashed curve shows events which are detected above baseline with $\Delta\chi^2>500$, and the dot-dashed curve shows those detected events which peak during an observing season. The solid grey lines show the theoretically expected asymptotic slope of the distribution, with power law slopes of $\pm 3$~\citep{Mao:1996mm}. The cyan data points show the timescale distribution observed by the MOA survey~\citep{Sumi:2011ffp}, which is uncorrected for detection efficiency and scaled arbitrarily -- this is most closely comparable to the dashed line showing all events detected by \euclid{}. The black lines and data points show the timescale distribution for events with detected planets. The solid line shows the timescale distribution of the host star microlensing event for $100$-$\mearth$ planet detections with no restriction on $\tzero$, while the data points show the same, but only for events where $\tzero$ lies in an observing season. The dashed and dot-dashed lines show the timescale distribution of detected $100$-$\mearth$ and $1$-$\mearth$ free-floating planet detections, respectively.}
\label{timescales}
\end{figure}

The primary observable of the microlensing lightcurve that is related to the host star's mass is the event timescale. The timescale is a degenerate combination of the total lens mass, the relative lens-source proper motion and the distances to the source and lens. Figure~\ref{timescales} plots the timescale distributions of all the microlensing events that occur within the observed fields and also the distributions for several cases of planet detections. The timescale distribution for bound planet detections is similar to the underlying timescale distribution, but is affected by the choice of detection criteria. Our third criterion, designed to select only events with well characterized timescales, cuts out potential detections in some long timescale events. Some of these events are detections of planets with large orbits, where the planetary lensing event is seen but the stellar host microlensing event is only partially covered (in which case the planet parameters may be poorly constrained) or may be missed completely (in which case the planet event would enter the free-floating planet sample). However, in other cases the cut on $\tzero$ is too zealous, and long timescale events with $\tzero$ outside the observing window, but with significant magnification in several seasons, are cut from the sample. The timescale of these events, and hence also the planetary parameters, are likely to be well constrained.

Figure~\ref{timescales} also plots the free-floating planet timescale distributions for planets of $1$ and $100$ Earth masses. Free-floating planets will dominate the timescale distribution at timescales less than a few days, if they exist in numbers similar to those suggested by \citet{Sumi:2011ffp}, which is twice the abundance that we have assumed. 

\begin{figure}
\includegraphics[width=\figuresize]{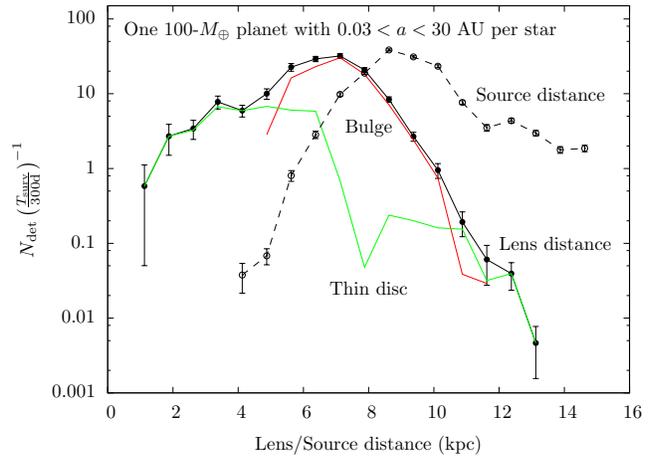}
\caption[Planet yield plotted against lens and source distance]{Predictions of the $100$-$\mearth$ planet yield as a function of lens (solid lines) and source (dashed line) distances, $\dl$ and $\ds$,
  respectively. The red and green lines show the contributions due to
  bulge and thin disc lenses, respectively; thick disc and halo lenses
  contribute the remainder, which is small.}
\label{NvDlDs}
\end{figure}

Figure~\ref{NvDlDs} plots the distribution of $100$-$\mearth$ planet
detections as a function of lens and source distances, $\dl$ and
$\ds$, respectively. The contribution of thin-disc and bulge
populations to
the yields is also plotted. Thick disc and stellar halo lens yields
have not been plotted as at no point are they dominant. However, near
the Galactic centre it should be noted that stellar halo lenses have a
higher yield than the thin disc due to the disc hole (see Section~\ref{galsim}). Most of the host stars are near-side bulge stars between
$5.5<\dl<8$~kpc. Beyond this, the number of lenses with detected
planets drops-off exponentially with increasing distance, dropping by four
orders of magnitude from $\dl \sim 9$ to $15$~kpc. The steepness of
this fall is partly caused by the truncation of the source
distribution at $15$~kpc. Though the majority of lenses are
in the bulge, a substantial number reside in the near disc. The
contribution of planet detections by each component is $68$, $27$, $1.2$
and $3.5$~percent for the bulge, thin disc, thick disc and stellar halo
populations, respectively. Unlike the lens stars, the majority of
source stars reside in the far bulge, with a small fraction in the far
disc. Very few near disc stars act as sources due to the low optical
depth to sources on the near side of the bulge.

\begin{figure}
\includegraphics[width=\figuresize]{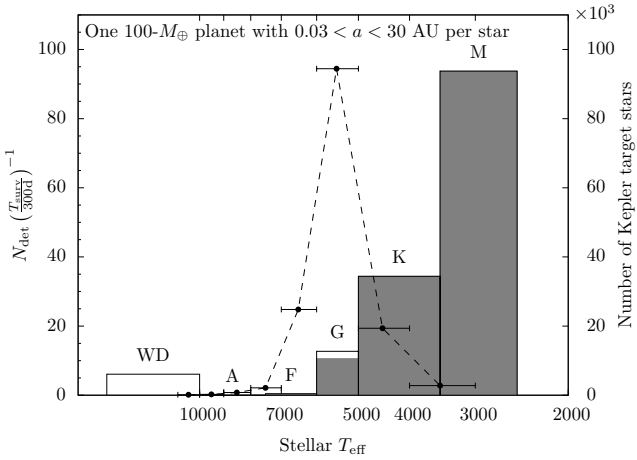}
\caption{Histogram of the number of $100$-$\mearth$ planet detections plotted against the effective temperature of the host star, binned according to the spectral type designations in the \besancon{} model. The shaded region shows the contribution due to main-sequence host stars, while white regions show the contribution of evolved host stars. The dotted line shows the distribution of effective temperatures of high-priority Kepler target stars~\citep{Batalha:2010}.}
\label{hostType}
\end{figure}

Figure~\ref{hostType} plots the distribution of $100$-$\mearth$ planet
detections as a function of the host star spectral type. The majority
of hosts are M dwarfs, but there are a significant number of
detections around G and K dwarfs and also white dwarfs. There will be
a negligible number of detections around F and earlier-type stars due
to their low number density. The distribution of planetary host stars probed by
\euclid{} is very different to that probed by any other technique. For
example, most of \euclid{}'s host stars are M dwarfs in the bulge, whereas
most of \kepler{}'s host stars are FGK dwarfs in the
disc~\citep{Howard:2012}. 

\section{Variations on the fiducial simulations} \label{variations}

In the previous section we have investigated the potential planet
yield of a \euclid{} microlensing survey and the properties of
detectable planets and their hosts. In this section we investigate how
the planet yield is affected by our choice of primary observing band,
the level of systematic photometry errors and the choice of spacecraft
design.

\subsection{Primary observing band} \label{bands}

We begin by examining the choice of primary observing band. The survey
strategy we have simulated involves the majority of observations being
taken in a primary band with a cadence of ${\sim} 18$~minutes while
auxiliary observations to gain colour information are taken every
${\sim} 12$~hours. We consider the use of each band available to
\euclid{}, $Y$, $J$ and $H$ in the near infrared using \nisp{} and
$RIZ$ using \vis{}. As \nisp{} and \vis{} can image the same field
concurrently we also consider simultaneous observations in $RIZ$ and
$H$. To maintain a comparable cadence, when $RIZ$ is the primary band
(or \vis{} is operating simultaneously with \nisp{}), the \vis{}
exposure times are $270$~s, as opposed to $540$~s when $RIZ$ is used as an
auxiliary band. In each scenario the total exposure time is identical, but the actual cadence is slightly different due to differences in the number of stacked images (we assume a $5$~s overhead between the images in the \nisp{} stacks, and the shutter on \vis{} takes $10$~s to open or close). As the sensitivities of the instruments in each band are slightly different, the images have different depths.

\begin{figure}
\includegraphics[width=\figuresize]{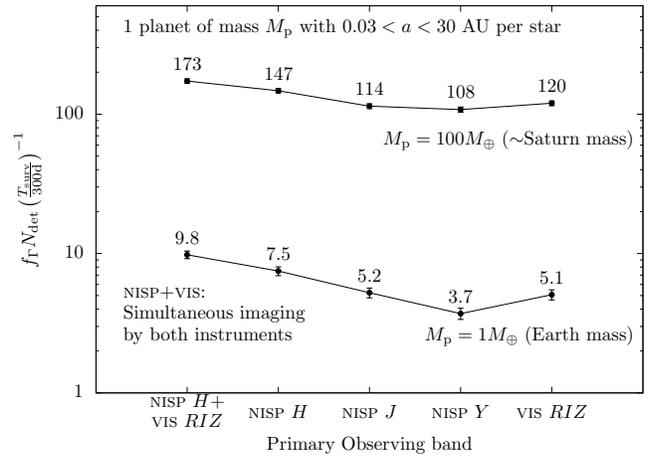}
\caption{Expected planet detections plotted against the different
  primary observing bands. Free-floating planet detections are not included.}
\label{Bands}
\end{figure}

Figure~\ref{Bands} shows the expected planet yields as a function of
the primary observing band. Focussing first on the scenarios without
simultaneous imaging, it is clear that $H$-band offers the highest
planet yields compared to the other two infrared bands. This is partly
due to the increased depth from a stack of $5$ images for $H$ as
opposed to a stack of $3$ images for $J$ and $Y$~\citep[the
  individual exposure times have been chosen optimize \euclid{}'s cosmological surveys;][]{redbook}. However, it is also due to the lower extinction
suffered in the $H$-band, and the correspondingly higher number
density of sources with magnitudes lower than the source catalogue
cut-off of $H_{\mathrm{vega}}<24$. 

The survey imaging with both available instruments
simultaneously obviously performs better than when using each instrument on its own. The
increase in yield is ${\sim}22\pm 4$~percent for both Saturn-mass and Earth-mass planets. As for the single primary instrument
scenarios, we require that the $\Delta\chi^2$ contribution of the
primary bands (the sum of $RIZ$ and $H$) to be greater than half the
total $\Delta\chi^2$. In reality, the expected yield of the
simultaneous imaging scenario represents an upper limit, as there are a
number of limitations that may preclude simultaneous imaging with
\vis{} for all pointings. These include losses due to cosmic rays,
which will affect ${\sim} 20$~percent of \vis{} data points \citep{redbook}, downlink bandwidth and power consumption limitations, which may only allow a
simultaneous \vis{} exposure every other pointing, say. The increase
in yield may therefore be small. However, the real value of
simultaneous \vis{} imaging will be the increased number of
exposures it is possible to stack in order to detect the lens
stars. This will greatly increase the depth of \vis{} images stacked
over the entire season, which in turn will allow the direct detection of more
lens stars, and hence an increase in the accuracy and number of mass measurements it is possible to make. We discuss this further in
Section~\ref{discuss}. Simultaneous \vis{} imaging will also allow source colours to be measured in many low-mass free-floating planet events, which will help to constrain their mass. It is clear therefore that as many simultaneous \vis{} exposures should be taken as possible.

\subsection{Systematic errors} \label{Sec:systematics}

There are many possible sources of systematic error, which can
include image reduction, photometry, image persistence in the
detector, scattered light, temperature changes in the telescope and
source and intrinsic variability in the source, lens or a blended star. The magnitude and
behaviour of each systematic will also be different; for example,
temperature changes will likely induce long-term trends in the
photometry, while image persistence may introduce a small
point-to-point scatter together with occasional, randomly-timed
outliers. It is likely that the systematics that produce long term
trends may be corrected for, to a large extent, either by using
additional spacecraft telemetry or by detrending similar to that used
in transiting exoplanet analyses~\citep[e.g.,][]{Holman:2010kms}. Even
for some systematics that behave more randomly, it may be possible to
account for and correct errors; for example, it may be possible to
correct for image persistence errors to a certain degree by using
preceding images. It is therefore difficult to predict the magnitude
and behaviour of systematics a priori. We therefore choose to model
systematic errors by assuming them to be Gaussian, and add the systematic component in quadrature to the standard photometric error. While likely a poor model for the actual systematics, it effectively introduces a floor below which it is not possible improve photometry by collecting more photons.

\begin{figure}
\includegraphics[width=\figuresize]{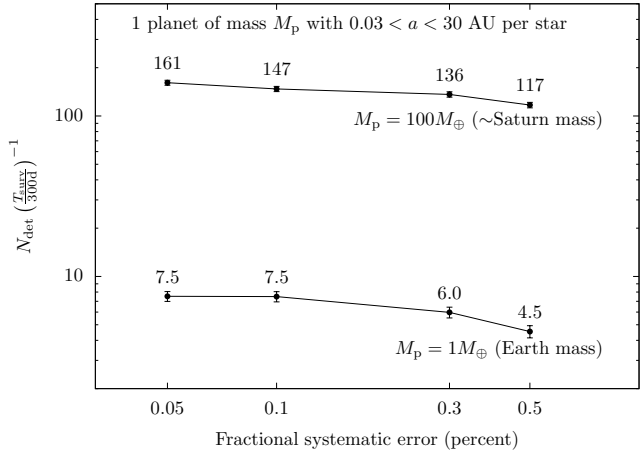}
\caption{Expected planet detections plotted against the size of the
  systematic error component.}
\label{Systematic}
\end{figure}

Figure~\ref{Systematic} plots the expected planet yield against
differing values of the systematic error component that we
assume. In all other simulations we have used the fiducial value of
the fractional systematic error
$\sigma_{\mathrm{sys}}=0.001$. Reducing the systematic error further
from this point does not provide a significant increase in yield, as
for the most part, at this level of systematic, photometric accuracy
is limited by photon noise. Increasing the systematic to
$\sigma_{\mathrm{sys}}=0.003$ does cause a drop in yields, by ${\sim}
8$ percent for giant planets to ${\sim} 20$ percent for
low-mass planets, as the systematic component becomes comparable to the photon noise. The situation is worse still for
$\sigma_{\mathrm{sys}}=0.005$, where the systematic component dominates. However, even with a systematic error
component this large and the conservative $\log$-$\log$ mass function,
${\sim} 4$--$5$ Earth-mass planet detections can be expected.

It is not possible at this stage to estimate the magnitude of
systematic error that should be used in our simulations, but it should
be noted that ground-based microlensing analyses often have systematic
errors of a similar magnitude to the values that we have
simulated~(D.~Bennett, private communication). The tight control of
systematics required by \euclid{} for galaxy-shape measurements should
mean that \euclid{}-\vis{} will be one of the best-characterized optical
instruments ever built \citep{redbook}; similarly, \nisp{} will
be optimized for performing accurate, photometry of faint
galaxies. Furthermore, \citet{Clanton:2012ird} recently showed that the HgCdTe detectors that will be used in \nisp{} can perform stable photometry to ${\sim} 50$ parts per million. How these considerations will relate to crowded-field photometry is not yet clear, but it is almost certain that the systematics will be lower than those achieved from the ground, potentially by a large factor. Our fiducial choice of a fractional systematic error $0.001$ ($1000$ parts per million) is therefore almost certainly conservative.

\subsection{Slewing time} \label{slewtime}

Another uncertainty in the yields we predict results from
uncertainties in the spacecraft design. 

Whilst the manufacturer and final design for the \euclid{} spacecraft is yet to be decided it is possible to 
explore some factors which are likely to have an important bearing on its microlensing survey capabilities.
One important factor is the choice of manoeuvring system used for slewing between fields. For fixed exposure and areal coverage the slew and settle time determines the cadence it is possible to
achieve on a particular field. Alternatively for larger slewing times one may shorten the exposure time to maintain cadence and areal coverage. Since the detection of low mass planets depends crucially on cadence this alternative approach is preferable when considering the impact of adjustments to the slewing time.

The slew time will ultimately depend on the technology used, in particular whether gas thrusters or reaction wheels are
used to perform field-to-field slews. A plausible range for the slew times based on initial design proposals is $85$--$285$~sec.

\begin{figure}
\includegraphics[width=\figuresize]{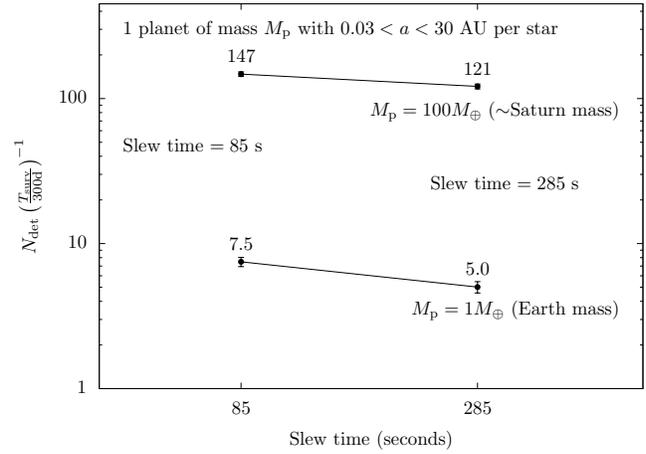}
\caption{Expected planet detections plotted for 85-sec and 285-sec slewing times, which encompass the likely range anticipated by different designs for the \euclid{} manoeuvring system.}
\label{Design}
\end{figure}

Figure~\ref{Design} shows the expected yield at either end of this slew time range. Maintaining a constant cadence of around 20~mins between repeat visits to a given field allows $270$~s per pointing of
stacked $H$-band exposure time for 85-sec slews and $108$~s for 285-sec slews. The increased depth allowed by a shorter slewing time produces a yield that is higher by $50\pm12$~percent at Earth mass and $22\pm5$~percent at Saturn
mass. 

\section{Summary discussion} \label{discuss}

The \euclid{} dark energy survey, which has been selected by ESA to fly in 2019, is likely to undertake additional legacy science programs. The design requirements of the \euclid{} weak lensing programme also make it very well suited to an exoplanet survey using microlensing and the \euclid{} Exoplanet Science Working Group has been set up to study this proposal. 

We have developed a baseline design for the Exoplanets \euclid{} Legacy Survey
(\exels{}) using a detailed simulation of microlensing.
The simulator, dubbed \mabuls{}, is based on the \besancon{} Galaxy model~\citep{Robin:2003bgm}. It is the first microlensing simulator to generate blending
and event parameter distributions in a self-consistent manner. and it enables
realistic comparisons of the performance of \euclid{} in different optical and infrared pass-bands.
We have used \mabuls{} to study a design for \exels{} with a total observing baseline of
up to $300$ days and a survey area of 1.6~deg$^2$. We show
that of the band-passes available to \euclid{} a survey
primarily conducted in $H$ will yield the largest number of planet
detections, with around 45 Earth-mass planets and even ${\sim}6$ Mars-mass planets. These numbers are based on current extrapolations of the exoplanet abundance determined by microlensing and radial velocity surveys. Such low-mass planets in the orbits probed by \euclid{} (all separations larger than $\sim 1$~AU) are presently 
inaccessible to any other planet detection technique, including
microlensing surveys from the ground.

While space-based
microlensing offers significantly higher yields per unit time than do
ground-based observations, this is not the only motivation for
space-based observations. A standard planetary microlensing event does
not automatically imply a measurement of planet mass or semimajor
axis, only the planet-star mass ratio and the projected star-planet
separation in units of the Einstein radius $\re$. To measure the
planet mass we must measure the lens mass, either by detecting subtle,
higher-order effects in the microlensing lightcurve, such as microlensing
parallax~\citep[e.g.,][]{Gould:2000nfm, An:2002eb5}, or directly detecting the
lens star~\citep{Alcock:2001ddl, Kozlowski:2007dd}. Without these the
mass can only be determined probabilistically~\citep[e.g.,][]{Dominik:2006spd,
  Beaulieu:2006fem}. The projected separation in physical units can be
determined if the lens mass and distance are known (as well as the
source distance, which it is possible to estimate from its colour and
magnitude). Determining the semimajor axis will require the detection
of orbital motion~\citep{Bennett:2010jsa,Skowron:2011kos}, but this
will only be possible in a subset of events~\citep{Penny:2011omm}.
For a survey by \euclid{} we expect parallax
measurements to be rare. Parallax effects are strongest in long
microlensing events lasting a substantial fraction of a year due to
the acceleration of the Earth~\citep{Gould:1992mss}, but \euclid{}'s
seasons will be too short to constrain or detect a parallax signal in
most events~\citep{Smith:2005pml}. 

However, thanks to the
high-resolution imaging capabilities of the \vis{} instrument, lens
detection should be common \citep{Bennett:2007phc}.  In events where the
light of the lens is detected, the lens mass and distance can be
determined by combining measurements of the angular Einstein radius
$\thetae$ (which gives a mass-distance relation) with a main-sequence
mass-luminosity relation. Measurement of $\thetae$ should be possible
for a large share of detected events, either from finite-source effects in the lightcurve or
by measuring the relative lens-source proper motion as the pair
separates~\citep{Bennett:2007phc}.
It is also possible to estimate the lens mass and distance from
measurements of its colour and magnitude~\citep{Bennett:2007phc}. From
a single epoch of \nisp{} and \vis{} images, this will likely not be
possible. However, over each 30-day observing period around $2000$ images will be
taken in \nisp{} $H$-band, with possibly a similar number with the \vis{} camera. These images will have random
pixel dither offsets. The images can therefore be stacked to form a much deeper, higher-resolution image in each band. From these images it should be possible
to isolate the source (whose brightness is known from the lightcurve)
from any blended light. After subtracting the source, if the remaining
light is due to the lens, its mass can be estimated from its colour
and magnitude. The planet mass can then be determined, as the
planet-host mass ratio is known from the lightcurve. However, if either the source or lens has a luminous
companion, estimating the lens mass will be more difficult~\citep{Bennett:2007phc}.

We have not attempted to estimate the number of planet detections with
mass measurements in the present work, but we aim to study this in a future work. 
These calculations will allow
a full determination of planetary microlensing figures of merit,
such as the one defined by the \wfirst{} Science Definition
Team~\citep{wfirstir}.

Finally, it is worth stating that our simulation of \exels{} has not been optimized. There are
many factors that can be varied to increase planet yields, such as the
choice of target fields, the number of target fields and the strategy
with which they are observed. However, planet yields are not the only
measure of the scientific yield of the survey. For example, planetary-mass measurements without the need for additional follow-up
observations would be an important goal of the \euclid{} microlensing
survey, and so any assessment of the relative performance of different
possible surveys must also evaluate performances in this
respect.

We have shown that \exels{} will be unrivalled in terms of its sensitivity to the cold exoplanet regime. A survey of at least six months total duration should be able to measure the exoplanet distribution function down to Earth mass over all host separations above 1~AU. This will fill in a major incompleteness in the current exoplanet discovery space which is vital for informing planet formation theories. This together with \exels{}'s ability to detect hot  exoplanets and sub-stellar objects (Paper~II) make it a very attractive addition to \euclid{}'s science capability.

\section*{Acknowledgements}

We thank Mark Cropper and Gregor Siedel for providing the \vis{} and \nisp{} PSFs. The computational element of this research was achieved using the High Throughput Computing facility of the Faculty of Engineering and Physical Sciences, The University of Manchester. MTP acknowledges the support of an STFC studentship. We are grateful to Scott Gaudi, Dave Bennett, David Nataf and Andy Gould for helpful discussions. We thank the anonymous referee, whose recommendations have improved the paper.

\newcommand{\jcap}{Journal of Cosmology and Astroparticle Physics}
\bibliographystyle{mn2e}


\end{document}